\documentclass[%
reprint,
amsmath,amssymb,  aps, prc, nofootinbib
]{revtex4-1}

\usepackage{graphicx}
\usepackage{dcolumn}
\usepackage{bm}

\begin{document}


\title{Shear viscosity of a hadron gas and influence of resonance lifetimes on relaxation
time}

\author{ J.-B.~Rose$^{1,2}$, J.~M.~Torres-Rincon$^{1}$, A.~Sch\"afer$^{1,2}$,
D.~R.~Oliinychenko$^{1}$, and H.~Petersen$^{1,2,3}$}

\affiliation{$^1$Frankfurt Institute for Advanced Studies, Ruth-Moufang-Strasse 1, 60438
Frankfurt am Main, Germany}
\affiliation{$^2$Institute for Theoretical Physics, Goethe University,
Max-von-Laue-Strasse 1, 60438 Frankfurt am Main, Germany}
\affiliation{$^3$GSI Helmholtzzentrum f\"ur Schwerionenforschung, Planckstr. 1, 64291
Darmstadt, Germany}


\date{\today}

\begin{abstract}
We address a discrepancy between different computations of $\eta/s$ (shear
viscosity over entropy density) of hadronic matter. Substantial deviations of this
coefficient are found between transport approaches mainly based on resonance propagation with finite lifetime
and other (semi-analytical) approaches with energy-dependent cross-sections, where interactions do
not introduce a timescale. We provide an independent extraction of this coefficient by using the
newly-developed SMASH (Simulating Many Accelerated Strongly interacting Hadrons) transport
code, which is an example of a mainly resonance-based approach. We compare the results from
SMASH with numerical solutions of the Boltzmann equation for simple systems using the Chapman-Enskog expansion, as well as previous results in the literature. Our conclusion is that the hadron
interaction via resonance formation/decay strongly affects the transport properties of the
system, resulting in significant differences in $\eta/s$ with respect to other approaches where
binary collisions dominate. We argue that the relaxation time of the system ---which
characterizes the shear viscosity--- is determined by the interplay between the mean-free
time and the lifetime of resonances. We show how an artificial shortening of the resonance
lifetimes, or the addition of a background elastic cross section nicely interpolate
between the two discrepant results.
\end{abstract}

\maketitle

\section{\label{intro}Introduction}

Studying physical properties of hot and dense nuclear matter is one of the main goals of
modern heavy-ion collision experiments. Among these properties, transport
coefficients are key elements as they control the non-equilibrium evolution of the
expanding fireball. One of the most well-studied dissipative
coefficients is the shear viscosity $\eta$, which measures the ability of the fluid system
to relax towards equilibrium after a shear
perturbation~\cite{lifschitz1983physical,de1980relativistic}. The interest in $\eta$
substantially increased after the realization that (almost) ideal fluid dynamics was able to describe the high elliptic flow that has been measured at 
the Relativistic Heavy-Ion Collider~\cite{Kolb:2000sd,Huovinen:2001cy}. In 2003, the first analytical computation of the shear viscosity over entropy density in a strongly coupled conformal gauge
theory using the anti-de Sitter/conformal field theory correspondence~\cite{Kovtun:2003wp,Kovtun:2004de} was performed. It was conjectured that $\eta/s=1/(4\pi)$ represents
a lower bound in any physical system, and in particular in Quantum
Chromodynamics (QCD), the theory of strong interactions governing the evolution of
a heavy ion collision. Extractions of the effective value of $\eta/s$ by fitting relativistic viscous hydrodynamics to experimental measurements have been carried out since then~\cite{Romatschke:2007mq,Dusling:2007gi} (see~\cite{Danielewicz:1984ww} for an early
estimate of $\eta$ in the context of heavy ion collisions). These results showed that the average $\eta/s$
in such systems is very close but slightly larger than the conjectured ratio. Some
attempts to study the temperature dependence $\eta/s(T)$ from experimental data were also
made within hydrodynamical~\cite{Denicol:2010tr,Niemi:2012ry} and hybrid
approaches~\cite{Song:2010aq,Song:2011qa}, and recent work has been done to extract it as part of a larger bayesian analysis of heavy ion collisions~\cite{Bernhard:2016tnd,Auvinen:2017fjw}

The low temperature behavior of the shear viscosity over entropy ratio can be constrained by calculations with hadronic degrees of freedom. For zero net baryon density, the shear viscosity of a hadron gas was studied up to temperatures of around $160$ MeV~\cite{Prakash:1993bt, Davesne:1995ms, Dobado:2003wr,
Muroya:2004pu, Chen:2006iga, Itakura:2007mx, Gorenstein:2007mw,
Dobado:2008vt,FernandezFraile:2009mi,Torres-Rincon:2012sda, Moroz:2013vd}, where the hadron gas turns into the
quark-gluon plasma in a crossover~\cite{Aoki:2006we}. Around the transition temperature,
results from gluodynamics and QCD on a lattice have provided estimates of $\eta/s$
~\cite{Meyer:2007ic,Meyer:2009jp,Mages:2015rea,Astrakhantsev:2017nrs}. In addition to the
temperature, the dependence of $\eta (T, \mu_B)$ on the baryon chemical
potential was also investigated ~\cite{Muroya:2004pu, Itakura:2007mx, Rougemont:2017tlu}.
It has been observed that the shear viscosity to entropy density ratio reaches a minimum around the phase
transition temperatures of everyday substances and several effective models of
QCD~\cite{Csernai:2006zz, Chen:2007jq, Dobado:2008vt, Sasaki:2008um,
Dobado:2009ek,Bluhm:2010qf,Torres-Rincon:2012sda}, which presents another motivation to
study this coefficient. It has been argued that $\eta/s$ has a minimum at
the critical point of a transition between the hadronic matter and the quark-gluon
plasma~\cite{Csernai:2006zz}, which is currently a subject of intense experimental
research.

Hadronic transport approaches aim to describe the effects of hadron rescattering in the
last stages of heavy ion collisions. One assumes that soon after hadronization the system is dilute enough
that it can be accurately described by a kinetic framework in terms of the Boltzmann
equation. Heavy ion collisions at low beam energies, where the production of a quark-gluon plasma is
unlikely, can also be appropriately described by such a model. In the latter scenario the
medium is dominated by hadrons at all times until the kinetic freeze-out, and the
transport approach covers the whole evolution of the system. Several transport codes have been developed to describe experimental observables in
heavy ion collisions~\cite{Ehehalt:1996uq, Bass:1998ca, Nara:1999dz,Cassing:2009vt,
Buss:2011mx,Novak:2013bqa}.

Our goal is to provide an independent computation of $\eta/s (T, \mu_B)$ in the hadronic
phase in the range of $T= 75-175$ MeV and $\mu_B = 0-600 $ MeV using SMASH (Simulating Many Accelerated Strongly-interacting Hadrons)\cite{Weil:2016zrk}. The
Green-Kubo relation~\cite{kubo,zubarev} is applied to thermalized hadronic matter in a static box
simulating infinite matter. Similar calculations have been performed in
~\cite{Muronga:2003tb,Demir:2008tr,Wesp:2011yy,Plumari:2012ep,Ozvenchuk:2012kh,Pratt:2016elw}.
The results of existing studies in this range disagree up to a factor of ten and our goal
is to understand the discrepancy between them. This question has
recently also been addressed in~\cite{Pratt:2016elw}, where the authors find a considerable
difference between the results from the UrQMD transport code~\cite{Demir:2008tr} (to which
SMASH is closer in conception), and the ones from the B3D transport approach
~\cite{Romatschke:2014gna,Pratt:2016elw}. To get a better understanding of the differences
between these approaches, we perform numerical solutions of the Boltzmann
equation for simple systems using the Chapman-Enskog expansion including genuine $2 \rightarrow 2$
collisions~\cite{Torres-Rincon:2012sda}. The main result of the present study is the explanation of the physical origin of this
discrepancy and, more generally, of the differences between transport computations whose
interactions are dominated by resonance formation, and those calculations in which binary
collisions dominate the dynamics.

In Sec.~\ref{method} we introduce the methodology to compute the shear viscosity of
infinite matter. In Sec.~\ref{kubo} we review the Green-Kubo technique to extract the
value of $\eta$ for an equilibrated system. Sec.~\ref{smash} presents an overview of the most relevant
features SMASH, the transport approach that we used. We describe the process of equilibration and
the extraction of thermodynamical quantities of the system in Sec.~\ref{entropy}. In Sec.~\ref{systematics}
we present a calculation in a simple system which allows us to study the systematic effects of the
parameters on the calculation. In Sec.~\ref{results}
we present our main results for $\eta$ in different hadronic systems. First, in
Sec.~\ref{pirho} we study a box with pions interacting via the formation of the $\rho$
resonance and compare it to a semi-analytical solution of the Boltzmann equation of a pure
pion gas. Then in Sec.~\ref{hadron_gas_results} we present $\eta/s$ and $\eta T/w$
(where $w$ is the enthalpy density) for the full hadron gas as a function of the
temperature and chemical potential. We compare our results with previous studies in
Sec.~\ref{discussion} and explain the origin of the main discrepancies between them.
Finally we present our conclusions and outlook in Sec.~\ref{conclusion}.

\section{\label{method}Methodology}

\subsection{\label{kubo}Green-Kubo formalism}

In this work we employ the Green-Kubo formalism for the shear viscosity calculation. More generally
this formalism describes how to relate transport coefficients to dissipative
fluxes, which are here understood as fluctuations around a state of equilibrium in a given
system~\cite{kubo,zubarev}. Specifically, assuming a uniform distribution of particles in space, the shear
viscosity $\eta$ is calculated using
\begin{equation}
  \eta = \frac{V}{T} \int_0^\infty dt \ C^{xy} (t),
  \label{kubo_shear_def}
\end{equation}
where $V$ is the volume of the system, $T$ its temperature, $t$ the time and
\begin{equation}
  C^{xy} (t) = \langle T^{xy} (0) T^{xy} (t) \rangle_{eq}
  \label{correlator}
\end{equation}
the auto-correlation function of off-diagonal components of the energy-momentum tensor
$T^{xy}$. In a transport approach such as SMASH, we have access to
the full phase space evolution of the system through knowledge of all point-like particles at all times. In a discrete case like this, $C^{xy} (t)$ takes the form
\begin{equation}
	C^{xy} (t) = \lim_{N \rightarrow \infty} \dfrac{1}{N} \ \sum_{s=0}^{N} T^{xy}(s \Delta t) \ T^{xy} (s \Delta t+t),
	\label{correlation_function}
\end{equation}
with $N$ being the total number of timesteps taken into consideration (in our calculation $N=5000$), and $\Delta t$ being the timestep size.
Note that we require the system to remain at thermal and chemical
equilibrium. Thus, $t=0$ is the onset of equilibrium.

The spatially-averaged energy-momentum tensor is defined according to
\begin{equation}
  T^{\mu\nu} = \frac{1}{V} \int d^3x \int d^3 p \frac{p^\mu p^\nu}{p^0} f({\bf x},{\bf p}) \ ,
  \label{tmunu_continuous}
\end{equation}
where $f({\bf x},{\bf p})$ is the phase-space density of the particles.
Discretized, this yields
\begin{equation}
  T^{\mu\nu} = \frac{1}{V} \sum_{i=1}^{N_{part}} \frac{p^\mu_i p^\nu_i}{p^0_i} \ ,
  \label{tmunu_discrete}
\end{equation}
where the sum is taken over all particles in the system, $N_{part}$.

It is generally thought that the auto-correlation function behaves as a decaying exponential~\cite{Demir:2008tr,Wesp:2011yy,Plumari:2012ep,Muronga:2003tb},
\begin{equation}
  C^{xy} (t) = C^{xy} (0) \ e^{-\frac{t}{\tau}} \ ,
  \label{correl_ansatz}
\end{equation}
where we introduce the shear relaxation time $\tau$. Using this ansatz, we find the
final expression that will be used to calculate the shear viscosity,
\begin{equation}
  \eta = \frac{C^{xy} (0) V \tau}{T} \ .
  \label{final_shear_eq}
\end{equation}
In a more detailed study of the systematic errors introduced in this method \cite{Rose:2017ntg}, it has been shown that there are signficant deviations from the exponential form at high densities ($\mu_B > 600$~MeV, $T > 175$~MeV). Therefore, the results in this work are restricted by this temperature-baryon chemical potential range. 

\begin{figure}
  \includegraphics[width=85mm]{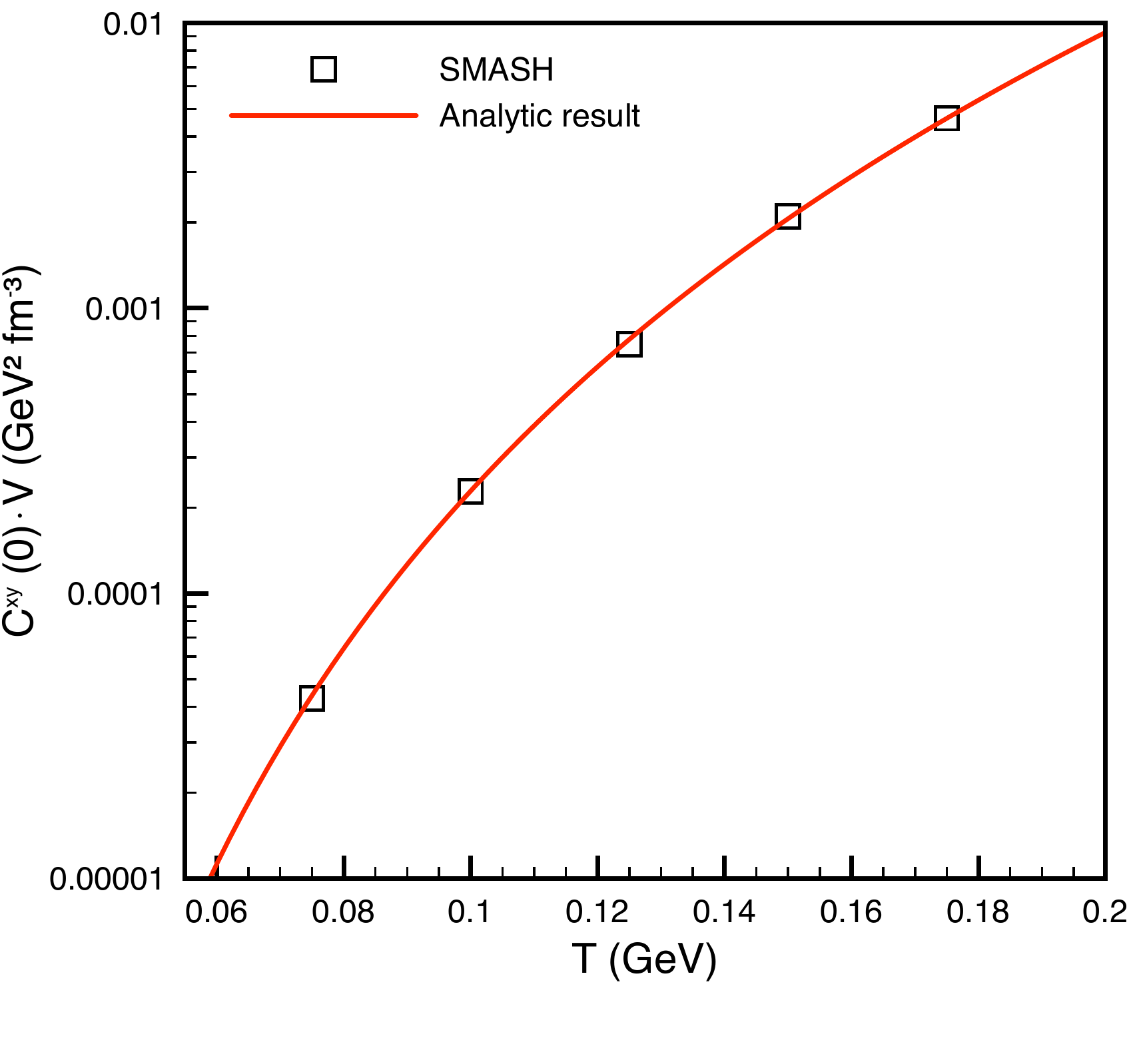}
  \caption{Volume-independent initial value of the correlation function as a function of temperature, for a system containing one species of particles of mass $m=138$ MeV.}
  \label{c0_vs_T}
\end{figure}

The initial value $C^{xy} (0)$ is also computed analytically after taking the continuum limit of
\begin{equation}
  C^{xy} (0) = \langle \sum_i \frac{(p_i^x)^2
  (p_i^y)^2}{V^2(p_i^0)^2} \rangle \rightarrow \int d^3{\bf x}
  d^3p \frac{ (p^x)^2 (p^y)^2}{ V^2(p^0)^2} f(p) \ ,
\end{equation}
with $f(p)$ being the Maxwell-Boltzmann distribution function. For a mixture of $N$ (stable)
hadrons
\begin{equation}
  C^{xy} (0)= \sum_{a=1}^N \frac{g_a z_a}{30 \pi^2 V} \int_0^\infty dp \
  \frac{p^6}{m_a^2+p^2} \exp \left( - \frac{\sqrt{m_a^2+p^2}}{T} \right) \ ,
  \label{cxy0}
\end{equation}
where $z_a=\exp(\mu_a/T)$ is the fugacity of the species $a$, with a spin-isopin degeneracy factor $g_a$.
Figure \ref{c0_vs_T} shows a comparison of the volume-independent $C^{xy} (0)$ for a single particle system as computed analytically and in SMASH.

One notes that the previous quantity does not depend on any parameter related to
the interaction of particles. Hence, all microscopical information about
the dynamics of the system (i.e. the cross-sections) is encapsulated within the
relaxation time of the correlation function. It can be interpreted
as the characteristic time for a fluctuation of $T^{xy}$ to decay, and is expected to
be of the order of the mean-free time (unless the cross-section is very forward-peaked). 

\begin{figure}
  \includegraphics[width=85mm]{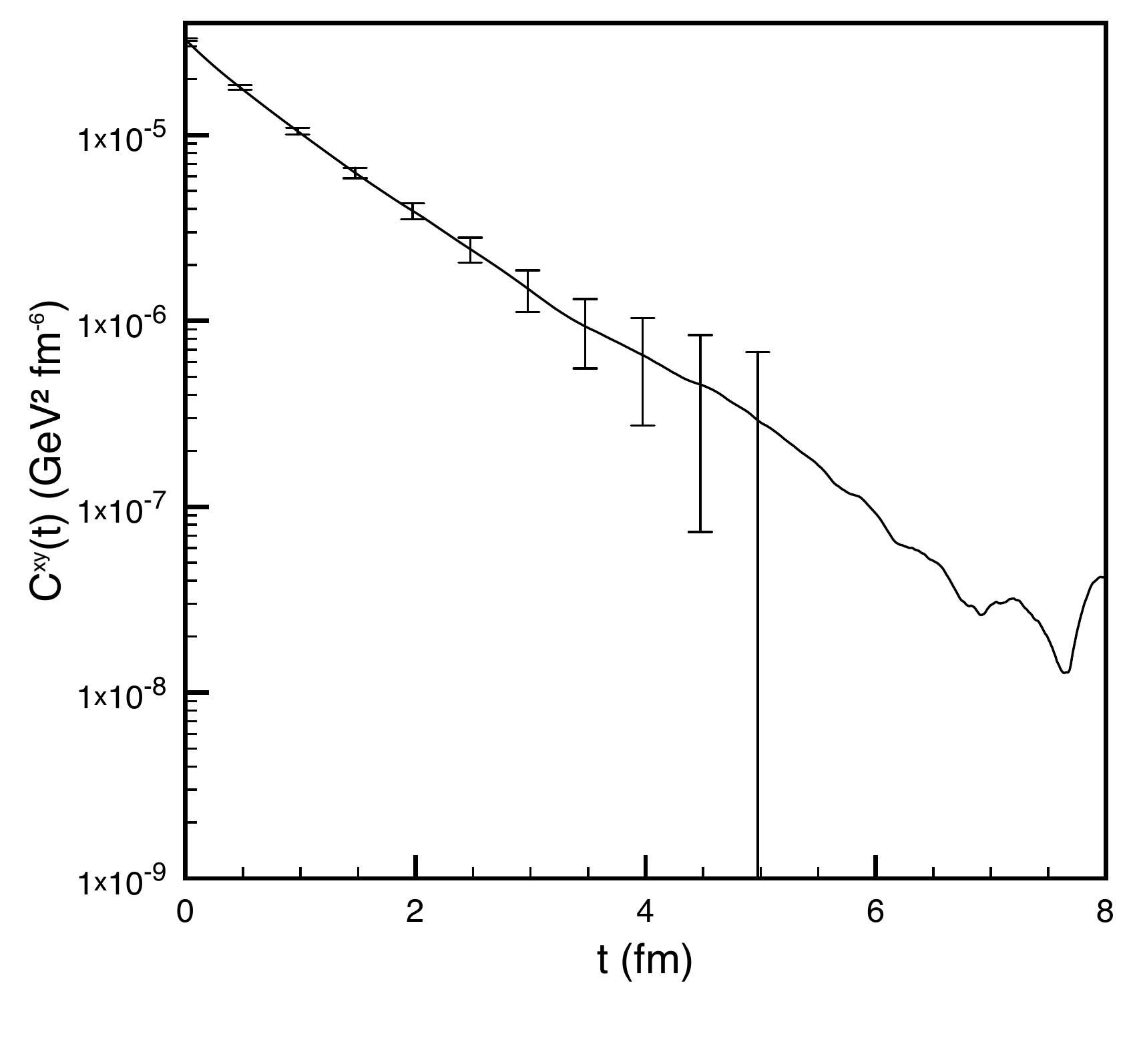}
  \caption{Typical example of a correlation function, for a massive gas of particles interacting through constant isotropic cross-sections $\sigma = 20$ mb, at temperature $T=200$ MeV and mass $m=138$ MeV.}
  \label{correl_example}
\end{figure}

The calculation in SMASH can in principle be performed by fitting the auto-correlation function to a decaying exponential according to  Eq.~(\ref{correl_ansatz}). This yields the parameters $C^{xy}
(0)$ and $\tau$. In practice however, problems arise with the upper limit of the sum in Eq.~(\ref{correlation_function}). 
As all simulations run for a
finite time, the sum over time intervals is performed over smaller and smaller data sets as the interval
grows. Hence, the error on the correlator rapidly increases with growing $t$ and becomes pure noise in the region of high $t$ (see Fig.~\ref{correl_example}). To cope with this problem we fit only the early part of the auto-correlation function (where the errors are still small) to a decaying exponential. Yet, 
such a procedure needs to be handled with caution, as it requires a proper determination of
the region to actually fit.

In this work, we use a fit method that relies on calculating the average auto-correlation
function from many simulations (typically 1000). The relative error is estimated for each
time interval $t$. As it is known that this error increases as $t$ increases (see
Fig.~\ref{correl_example}), we implement a criterion for the cutoff to happen when the
relative error on the average auto-correlation function reaches a given level. By looking at systems of varying complexity we show in Ref.~\cite{Rose:2017ntg} that one should use cutoff values between 4-10\% .
For all further calculations, a cutoff of 6\% was chosen to fit the average correlation function.

\subsection{\label{smash}Hadronic transport: SMASH}

SMASH~\cite{Weil:2016zrk} is a recently developed transport approach to describe the hadronic evolution within heavy-ion collisions at different accelerators like the Large Hadron Collider (LHC), the Relativistic Heavy Ion Collider (RHIC) and the SIS-18 at the GSI Helmholtzzentrum f\"ur
Schwerionenforschung. Effectively, a set of coupled Boltzmann
equations~\cite{de1980relativistic} for different hadron species,
\begin{equation} \label{eq:boltzmann}
 p^\mu \frac{ \partial f_i (t,{\bf x},{\bf p})}{\partial x^\mu}  = C_{coll} [f_i,f_j] \ ,
\end{equation}
where $f_i$ denotes the one-particle distribution function defined in the phase space of
the species $i$, is solved numerically. $C_{coll}$ is the collision integral involving all the distribution functions of the other species which interact with $i$.

The particles evolve according to their equations of
motion and are allowed to collide. SMASH uses a geometrical collision criterion based on the total cross section (see \cite{Weil:2016zrk} for
details). The degrees of freedom in \cite{Weil:2016zrk} have been expanded to contain all the well established hadrons listed in \cite{Olive:2016xmw}, except the $\pi_2(1880)$ and $\phi(2170)$ light unflavoured mesons and the $N(1990)$, $N(2600)$ and $\Delta (2420)$ baryons.
It is also important to mention that at low energies the interactions among hadrons
are assumed to happen via resonance formation (this is supposed to be valid up to $\sqrt{s}$ of
several GeV, where resonant structures disappear from hadron-hadron cross sections).
Therefore, in the standard application of SMASH, almost all reactions are of the type $2
\rightarrow 1$ and $1 \rightarrow 2$ (with the notable exception of the nucleon-nucleon
interaction, where the cross section is introduced as a parametrization of the experimental
data).

SMASH is able to mimic an infinite medium by running a ``box
calculation'', where interactions can be chosen to be either 1) purely elastic with
constant $2\rightarrow 2$ cross sections, 2) based on the previously described resonance
model, or 3) a combination of both. 

Another issue to consider is the baryon-antibaryon annihilation, which are usually treated by string fragmentation ~\cite{Ehehalt:1996uq, Bass:1998ca, Nara:1999dz,Cassing:2009vt,
Buss:2011mx,Novak:2013bqa}. This is however problematic in infinite matter
calculations, as detailed balance is not conserved for string fragmentation processes. In other words, while it is possible to annihilate a baryon and an
anti-baryon to produce many particles, the reverse process is computationally challenging. One possible solution to this problem is
the production of resonances instead of strings in such scatterings. Using an appropriately
parametrized cross-section, we rely on the fact that on average, nucleon-antinucleon
annihilation produces $5$ pions, as it was suggested in \cite{Demir2010Thesis}. Thus we implement the following reaction:
\begin{equation}
  \bar NN \leftrightarrow h_1 (1170) \rho
\end{equation}
In SMASH, the $\rho$ resonance decays exclusively to $2\pi$ (when neglecting
electromagnetic interactions) and the $h_1 (1170)$ resonance decays to $\pi \rho$. We then
get, after resonance decays, 5 pions from every $\bar NN$ interaction. This process is
reversible in all steps and we recover detailed balance for nucleon-antinucleon annihilation.

\subsection{\label{entropy}Thermodynamic quantities}

In complex systems where inelastic collisions are allowed, the chemical composition
of the system, its temperature and the chemical potential can change from the initial state if this one
is not directly equilibrated. One thus needs a way to calculate the 
actual values of these thermodynamic quantities in the system after equilibration.

The temperature is obtained by fitting momentum distributions of given particle
species:
\begin{equation}
  \frac{dN}{dp} \propto p^2 e^{-\frac{\sqrt{p^2+m^2} - \mu}{T}} \ .
  \label{temperature_fit}
\end{equation}
Note that the extracted temperatures differ slightly from
one species to the next. It is therefore necessary to distinguish between the
temperature of a particle species and the temperature of the system. In concrete terms, we
will consider the temperature of the system to be the weighted average of the most
abundant stable particles in any system (in the case of the full hadron gas described in
section \ref{hadron_gas_results}, this will typically be pions, kaons and nucleons, where their respective multiplicities are taken as weights).

Although there is in theory a different chemical potential for every particle species, we will here only be interested
in true conserved quantum numbers; specifically, the baryon chemical potential. It is assumed that
the chemical potential of baryons can be approximated by that of nucleons.
The latter is obtained by using the ratio of the momentum distributions (Eq.~\ref{temperature_fit})
of nucleons to that of anti-nucleons, such that

\begin{equation}
  \frac{dN_N/dp}{dN_{\bar N}/dp} = \exp{\left( \frac{2\mu_B}{T} \right)} \ .
  \label{muB_fit}
\end{equation}
This ratio is approximately flat in the region which was used for the temperature determination.
Its momentum average in this region is calculated and used as a proxy for the baryon chemical potential.

Finally, let us mention that we use the definition of the Gibbs free energy to
calculate the entropy density,

\begin{equation}
  s = \frac{w-\mu_B n_B}{T} = \frac{\epsilon + P - \mu_B n_B}{T}
  \label{gibbs_entropy}
\end{equation}

where we introduce the enthalpy $w$, energy density $\epsilon$ and pressure $P$.
$\epsilon$ and $P$ are obtained directly from the diagonal components of the averaged 
energy-momentum tensor, the temperature $T$ and
baryon chemical potential $\mu_B$ from Eqs.~(\ref{temperature_fit}) and (\ref{muB_fit})
and the baryon number density $n_B$ by counting baryons and anti-baryons in a given
volume of the system.

\begin{figure}
  \includegraphics[width=85mm]{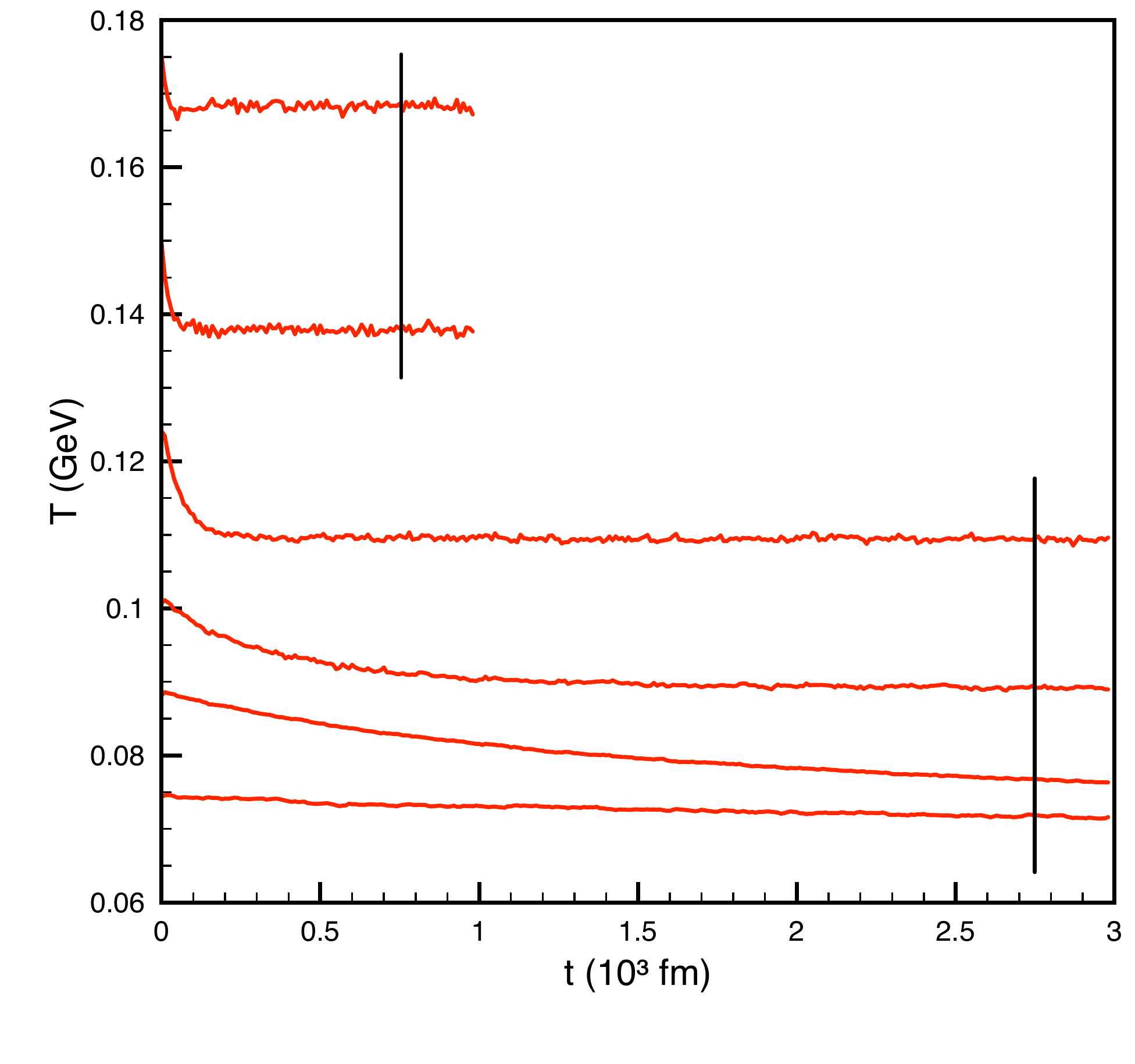}
  \caption{Temperature evolution of the box over time, for different initial temperatures. The vertical bar shows which part of the evolution is considered as in equilbrium and taken into account in the calculation of the correlation function. Note that the region of thermal equilibrium is in some cases much larger than the considered one.}
  \label{temp_vs_t}
\end{figure}

As mentioned in section \ref{kubo}, the system is required to be in thermal and chemical equilibrium for the Green-Kubo formalism to be applicable.  To ensure that such an equilibrium is reached fast, every particle species
density (including resonances) is initialized near the thermal expectation (using Boltzmann
statistics), allowing for Poissonian fluctuations, following
\begin{equation}
  n_a = \frac{g_aT^3 e^{\mu_a/T}}{2\pi^2} \frac{m_a^2}{T^2} K_2 \left(\frac{m_a}{T}\right),
  \label{grand_canon}
\end{equation}
where $a$ is a given particle species.
Transport model densities in such
calculations typically equilibrate to values near the Boltzmann grand canonical expectation. 
Small deviations are expected since the initialization of particle multiplicities does not take into account
resonance spectral functions, which leads to slightly different final particle densities.
The infinite matter simulation is thus left to equilibrate for an
appropriate time, and the viscosity calculation proceeds after both chemical and
thermal equilibrium have been reached. The chemical equilibrium is checked by verifying that
the multiplicities of the individual species in the box saturate to a stable value
(see~\cite{Weil:2016zrk} for examples). Similarly, the thermal equilibrium requirement is checked by monitoring the temperature of the box and waiting for it to reach a saturation value. 
Thermal equilibration takes much longer than chemical equilibration.

Equilibration times depend strongly on the complexity of the content of the box, more degrees of freedom corresponding to longer equilibration times. In the trivial case presented in the next section, only one species of particles is allowed to interact elastically. In this setup, the system is directly initialized
into chemical and thermal equilibrium, since no particle number changing processes can occur.
For the full hadron gas, the equilibration process however lasts markedly longer, especially at low
temperatures (which is expected, as initial density increases fast as a function of temperature, see Eq.~(\ref{grand_canon})). As seen in Fig.~\ref{temp_vs_t}, thermal equilibration times for such a system
usually range from a couple hundred fm at higher temperatures ($T=150$ MeV and higher)
to several thousand fm at lower temperatures ($T=100$ MeV and lower).

\subsection{\label{systematics}Systematics}

One of the simplest hadron gas systems that one can think of is composed of one species of
particles that only interacts elastically via a constant isotropic cross-section. These
systems have been studied extensively and their shear viscosity can be extracted
analytically by linearizing the collision term of the Boltzmann equation using the
Chapman-Enskog or relaxation time approximations~\cite{Plumari:2012ep}. As such, this
system constitutes the perfect playground for a proof of concept. The main goal of this
first study is the evaluation of the systematic error of the
present calculation by comparing it to a well-known and understood case.

A numerical solution of the Boltzmann equation is obtained following the 
methodology of \cite{Torres-Rincon:2012sda}. We implement the Chapman-Enskog expansion~\cite{de1980relativistic}
to the (nonequilibrium) distribution function in the Boltzmann equation~(\ref{eq:boltzmann}). 
This approach, consistent with the hydrodynamic expansion, allows us to linearize the collision operator and simplify the left-hand side of this 
equation by replacing the distribution function by the local equilibrium distribution.
After expanding the deviation from equilibrium in an appropriate polynomial basis, we transform the 
integral Boltzmann equation into a matricial equation which is solved order by order in the polynomial expansion.
Matching the microscopic expression of the energy momentum tensor~(Eq.~\ref{tmunu_continuous}) with the Newton equation 
(in the local rest frame, $u^\mu =(1,0,0,0)$)
\begin{equation}
  T^{xy} = -\eta \left( \frac{\partial u^x ({\bf x})}{\partial y} + \frac{\partial u^y ({\bf x})}{\partial x} \right) \ , 
  \end{equation}
the value of $\eta$ is extracted due to small deviations from equilibrium.

\begin{figure*}
\centering
 \includegraphics[width=59mm]{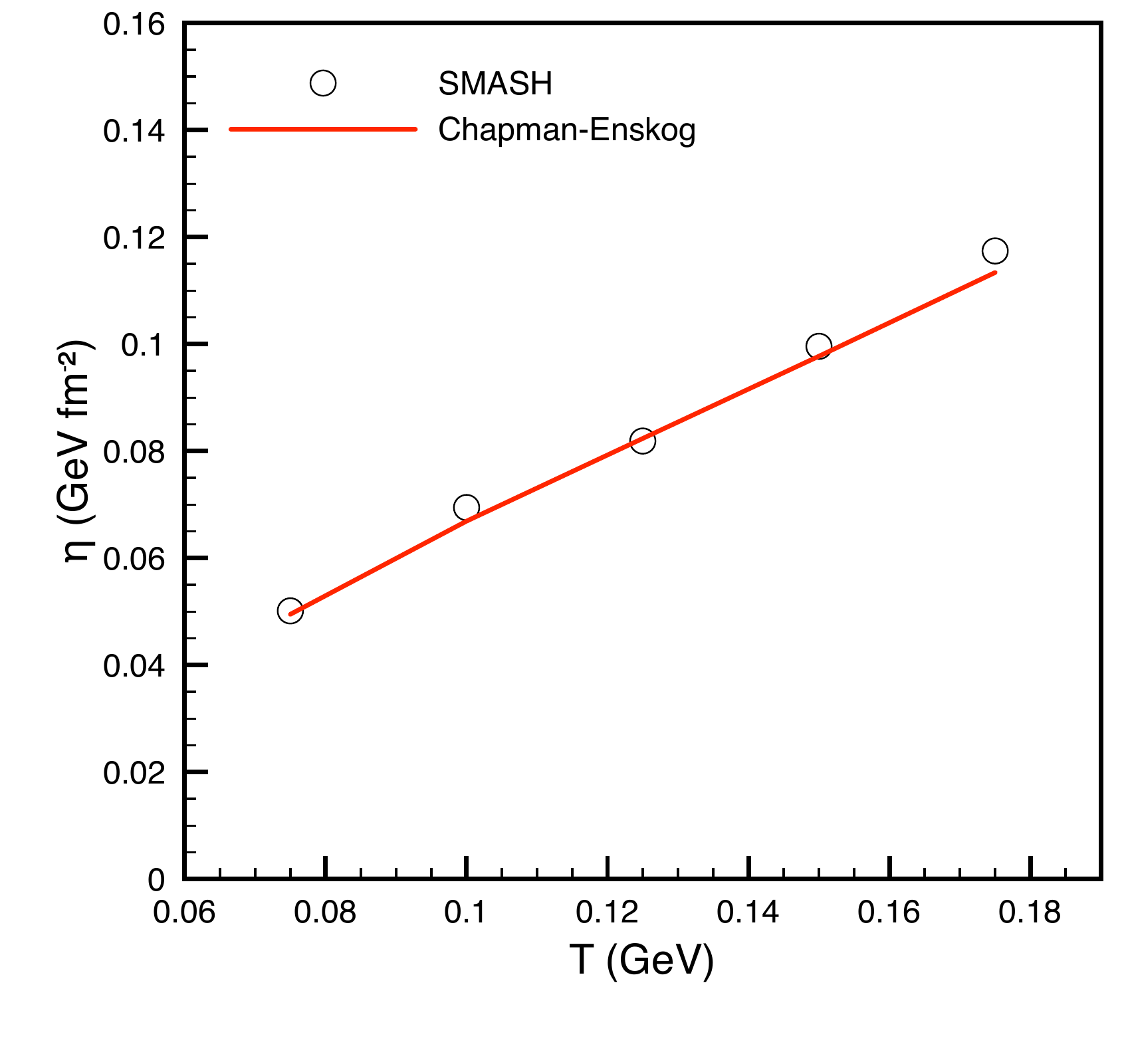}
 \includegraphics[width=59mm]{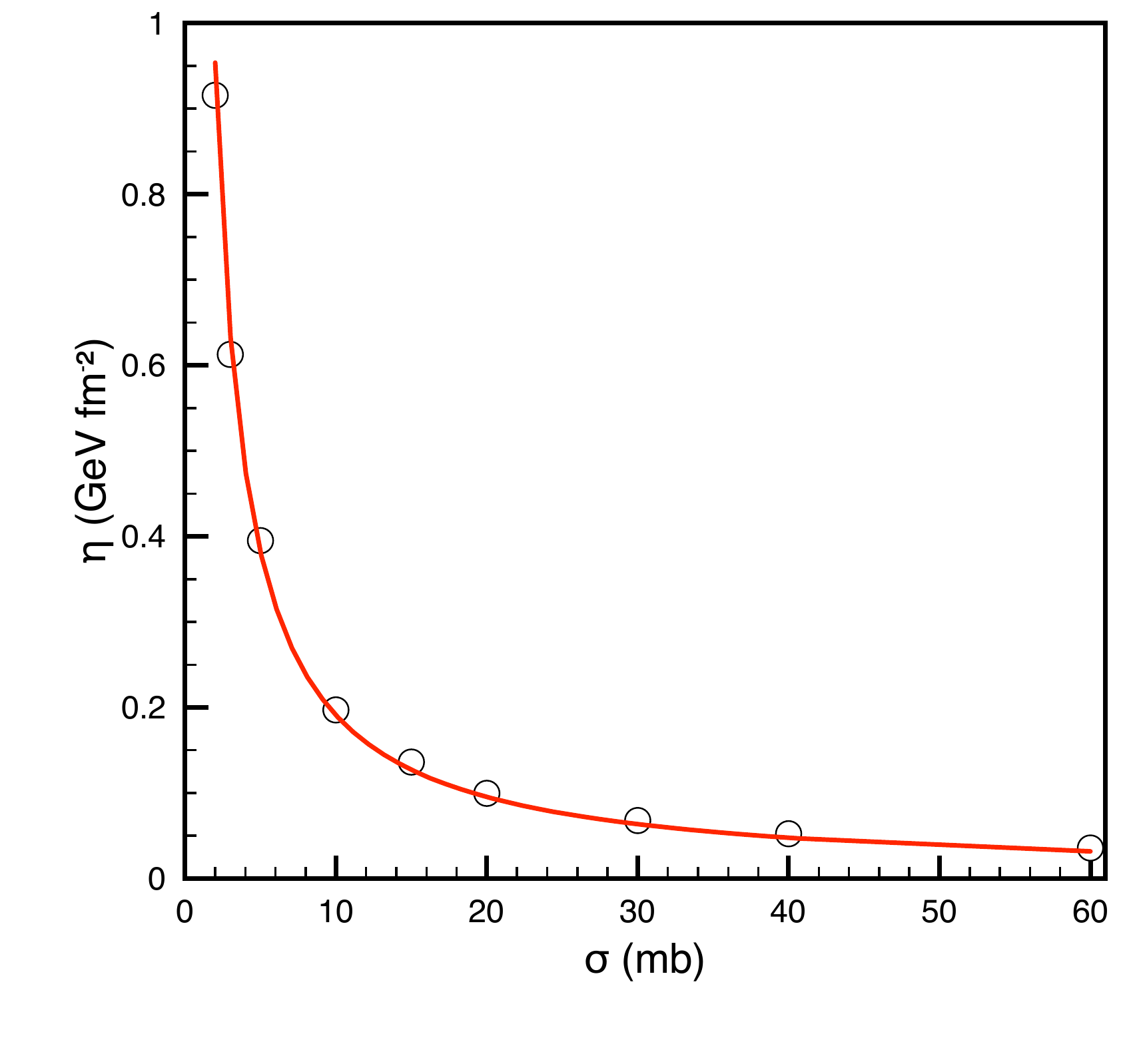}
 \includegraphics[width=59mm]{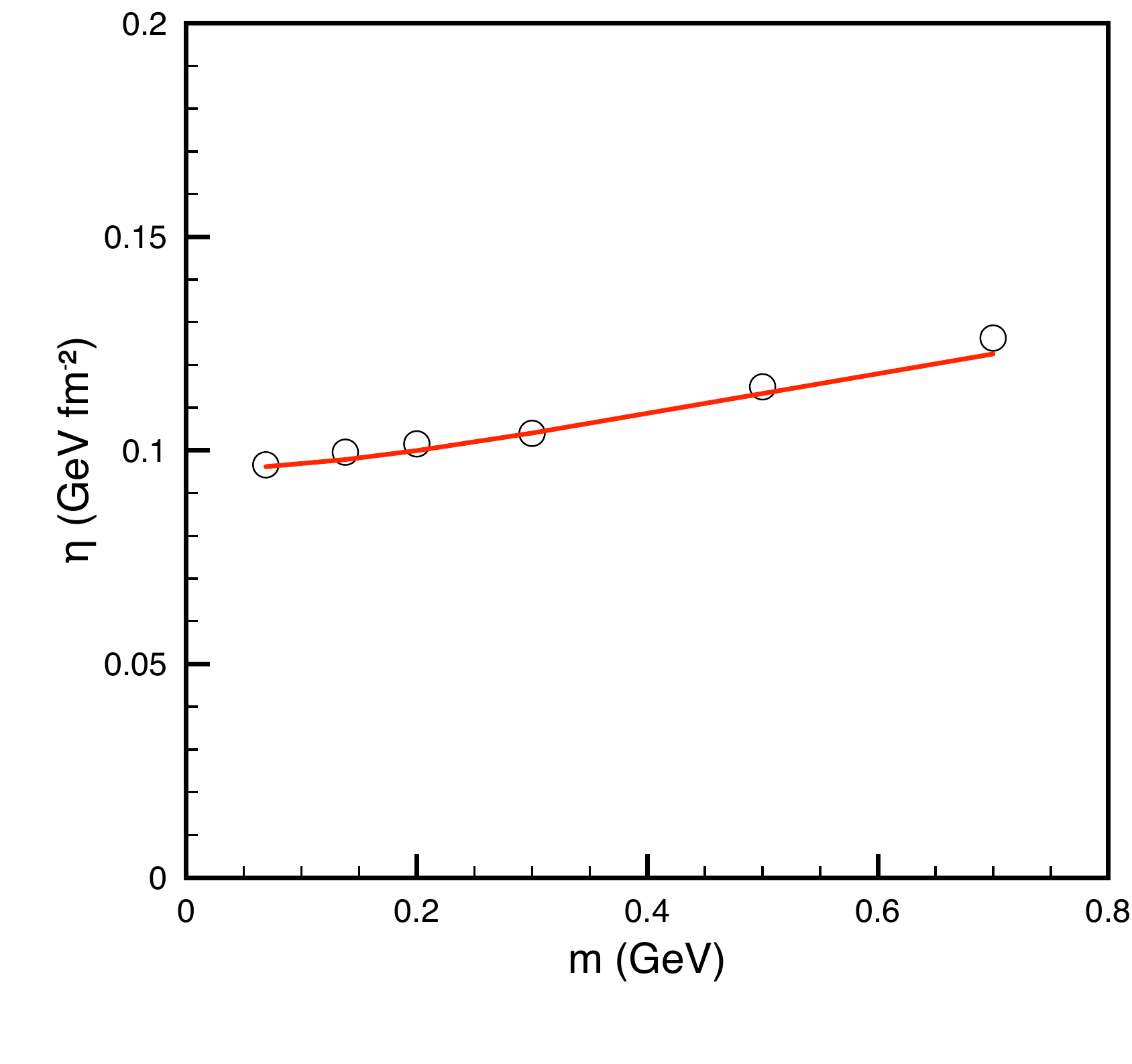}
\hspace{0mm}
 \includegraphics[width=59mm]{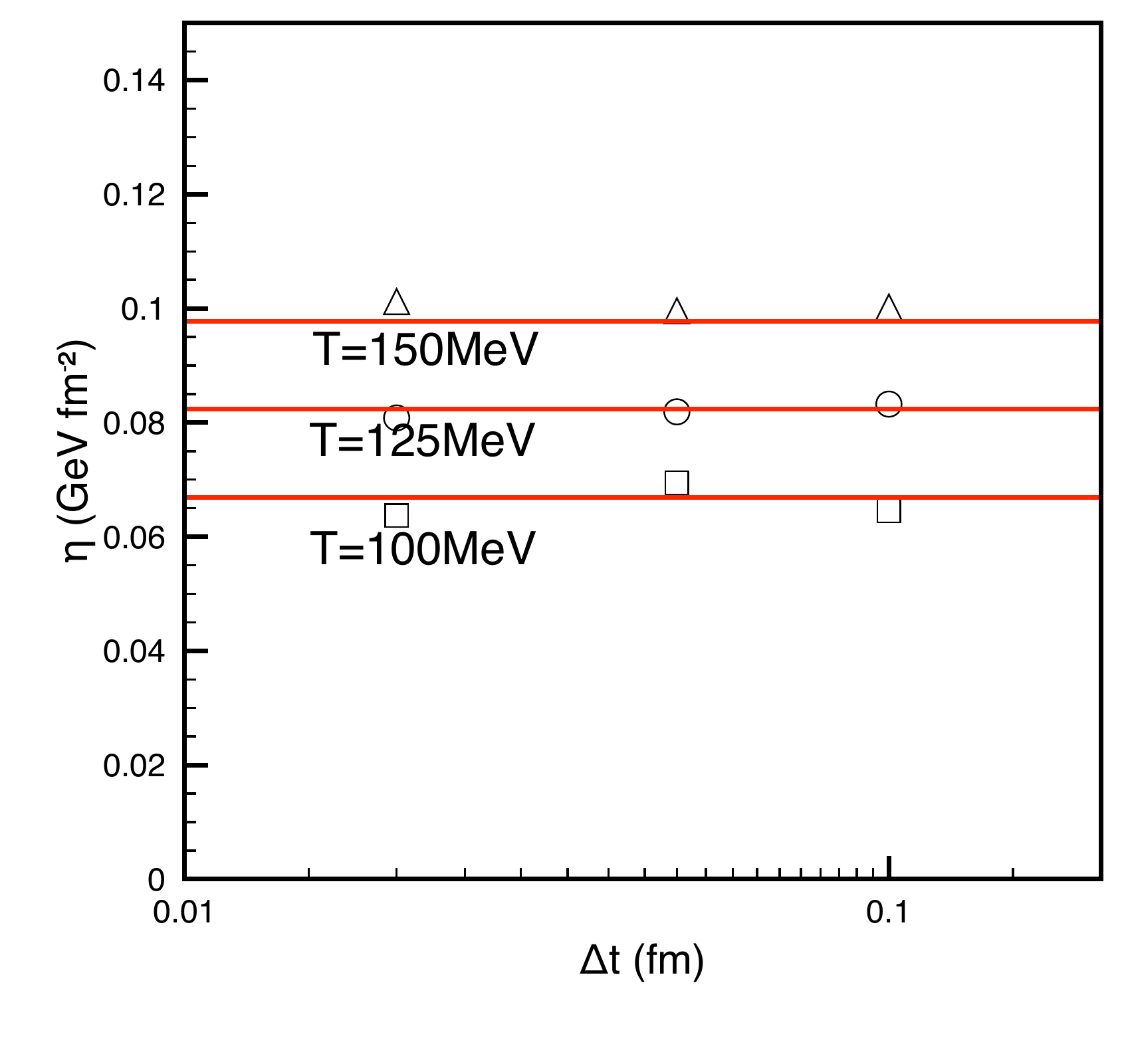}
 \includegraphics[width=59mm]{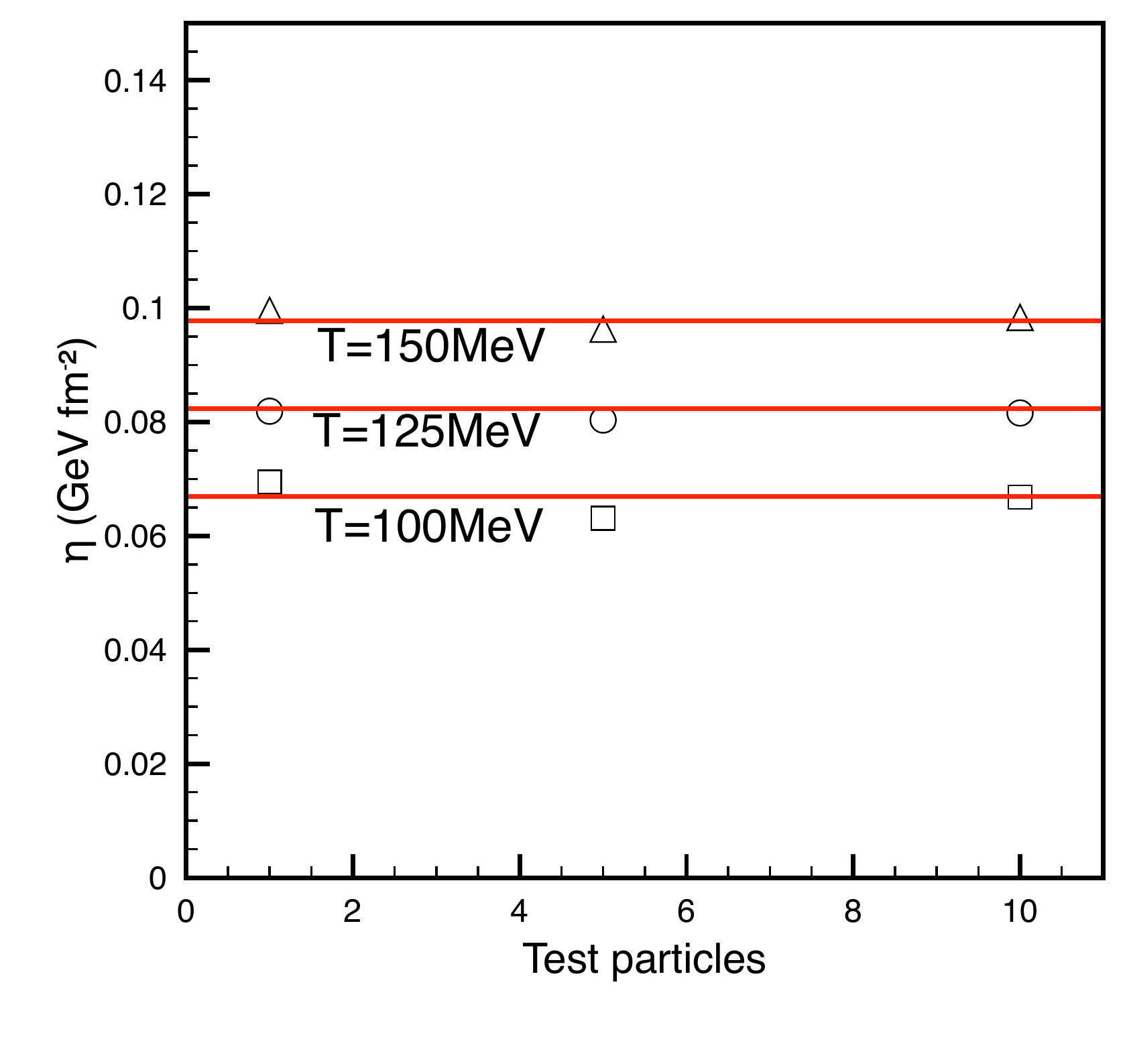}
 \includegraphics[width=59mm]{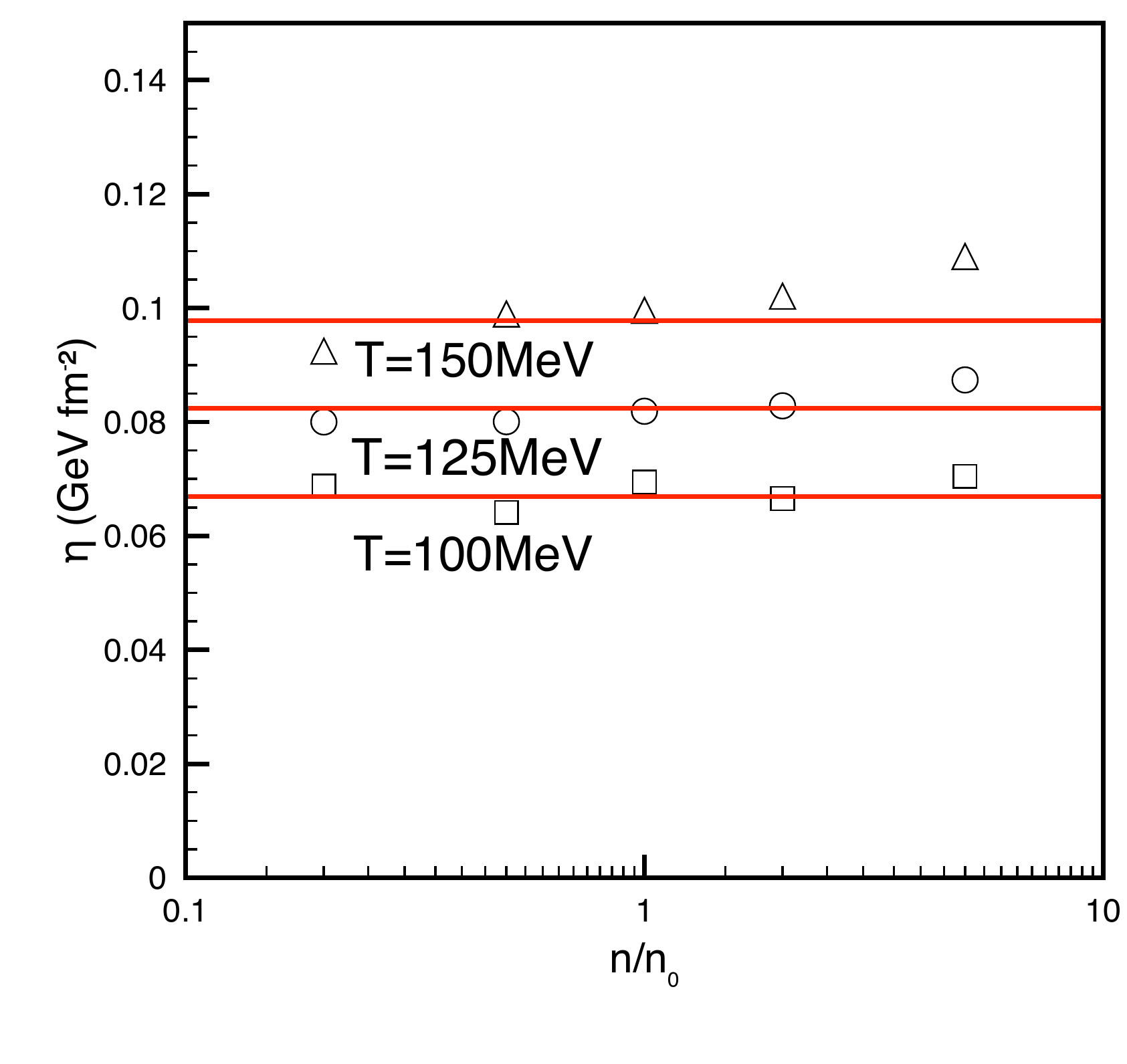}
\caption{Single-species gas systematics. Viscosity of a massive gas interacting through constant elastic cross-section, as computed in SMASH (symbols) and a Chapman-Enskog approach (red lines), plotted versus temperature (top left), cross-section (top middle), mass of the particle (top right), time step size between calculations of $T^{xy}$ (bottom left), number of test particles (bottom middle) and density ratio $n/n_0$ (bottom right), where $n_0$ is the particle density at zero chemical potential. When not mentioned directly on the plot or on an axis, $T=0.15$ GeV, $\sigma=20$ mb and $m=0.138$ GeV.}
\label{pion_sys}
\end{figure*}

Fig. \ref{pion_sys} shows the result of a comparison between the SMASH infinite matter calculation employing the Green-Kubo formalism to a 15th order Chapman-Enskog
calculation~\cite{Torres-Rincon:2012sda}. As is readily apparent, the previously described
way to fit the Kubo method reproduces analytical calculations very precisely.
The top three panels show the dependence of the shear viscosity on the three physical
parameters that appear in this calculation, namely the temperature, the constant elastic
cross-section and the mass of the particles, all of these being otherwise kept equal. We remind
the reader that the kinetic theory estimates of $\eta$ for a system of ultrarelativistic
particles interacting with a constant cross section is $\eta \sim
T/\sigma$~\cite{de1980relativistic,Plumari:2012ep} and for nonrelativistic particles is
$\eta \sim (Tm)^{1/2}/\sigma$~\cite{Torres-Rincon:2012sda}. We observe that the shear
viscosity increases with temperature and mass, and decreases with cross-section. This behavior is expected, since the relaxation time to equilibrium decreases
as the cross section and thus the number of collisions increases. The dependence on the cross section is very well approximated  by
$1/\sigma$, while the precise scaling with $T$ and $m$ follows an intermediate behavior
between the nonrelativistic and ultrarelativistic cases.

The three bottom panels of Fig.~\ref{pion_sys} refer to the method's dependence on
more technical parameters. The one on the left shows that, provided the use of a
sufficiently small timestep size, the result converges to the analytical value. We find that the range of timestep sizes considered is appropriate; all further calculations use a timestep size of 0.05 fm.
The bottom middle plot shows
the effect of including test particles. In this case each physical particle is divided into
multiple ones while correspondingly scaling down each components' cross-section, thus
approaching the continuum limit. Very limited effects are observed. Hence, for
simplicity, and since the use of many test particles implies heavy computational costs, all
further calculations are made using only one physical particle. Note that this result differs from what is found in the literature~\cite{Plumari:2012ep},
where the use of hundreds of test particles is recommended. Since $\tau$ is independent from $N_{test}$ in a local collision kernel, it follows that the non-locality of the geometrical collision criterion could explain differences in viscosity from the number of test particles~\cite{Cheng:2001dz}. The apparent
discrepancy can be explained by the fact that we use similarly large numbers of box
calculations instead of test particles for computational convenience. The last plot, to the bottom
right, explores the effect of altering the density of the system. In principle, it is
well known that the shear viscosity is independent of the density~\cite{Maxwell}.
Within our calculation however, it is possible that numerics have an effect on observables in
some limits. Yet, as the last panel shows, these effects prove to be negligible in most
cases, with the exception of very large densities at higher temperatures. In
any case, these non-zero deviations remain small with respect to the value of the
analytical calculation.

This first test scenario shows that, as expected, the results of the method are mostly unaffected by the variation of non-physical parameters. Thus it is applicable in a
wide range of more complex situations. The maximum deviation from analytical calculations is observed to be less than 8\%. Therefore, this value is assigned to be our systematic error in all further calculations.

\section{\label{results}Results}

Now that a firm basis for the calculation has been established, we use it in a succession
of systems of increasing complexity.

\subsection{\label{pirho}$\pi-\rho$ system}

\begin{figure}
  \includegraphics[width=70mm]{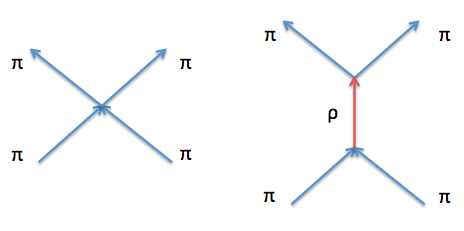}
  \caption{Sketch of the point-like vs resonant picture of interactions.}
  \label{lifetime_vs_point}
\end{figure}

The first system consists of pions interacting through a $\rho$ resonance which is
described by a Breit-Wigner mass distribution. In SMASH, the pion pair is produced isotropically.

Analytical calculations of the viscosity of such systems using the Chapman-Enskog
formalism exist~\cite{Prakash:1993bt,Davesne:1995ms,Torres-Rincon:2012sda}. These analytical calculations
consider a system of pions interacting via a cross-section that reproduces the $\rho$
peak, but the resonance is actually never produced, the outgoing pions being
directly created in a point-like interaction. Figure \ref{lifetime_vs_point} illustrates the
schematic difference between the two descriptions; as one can see, the main difference between the
two pictures is the fact that in SMASH, the $\rho$ resonance has a finite non-zero
lifetime.

\begin{figure}
  \includegraphics[width=85mm]{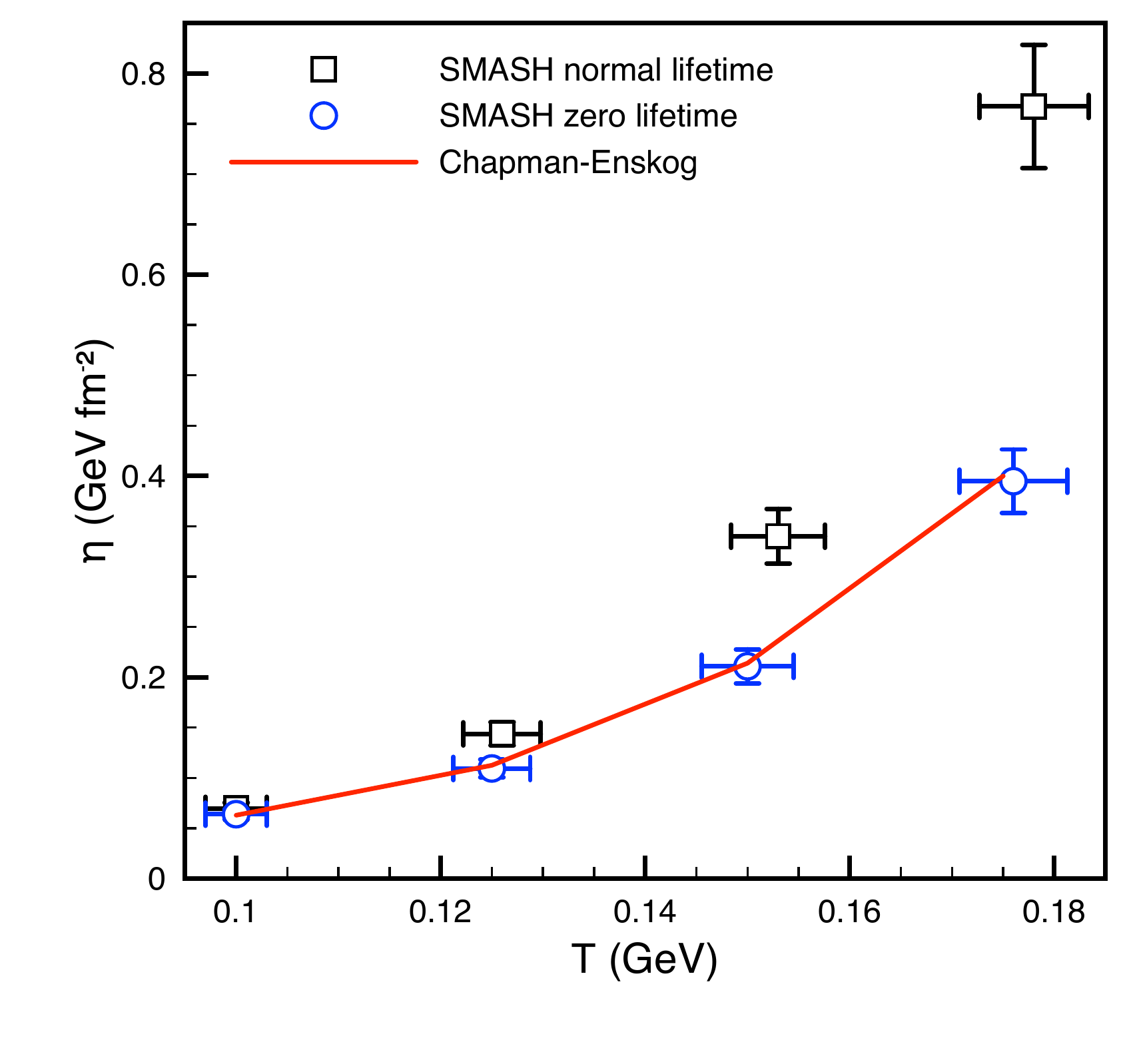}
  \caption{Shear viscosity vs temperature in a $\pi-\rho$ gas, for different lifetimes and compared to the Chapman-Enskog analytical result.}
  \label{pirho_viscosity}
\end{figure}

For the sake of comparison, several modifications have been
made in the approach presented in \cite{Torres-Rincon:2012sda}: 1) only the $(I,J)=(1,1)$ channel (relevant for the $\rho$ meson) is kept in the 
$\pi\pi$ scattering, whereas the isoscalar and isotensor channels are neglected, 2) in spite of the genuine $p$-wave 
scattering in the isovector channel, the differential cross section is tuned to be isotropic for consistency with SMASH~\footnote{The shear
viscosity is inversely proportional to the ``transport cross section'',
$\sigma_{tr}(s)=\int d\Omega \sin^2 \theta \ d\sigma/d\Omega (s,\theta)$~\cite{Plumari:2012ep}. For an $s-$wave isotropic interaction one has $\sigma_{tr}=2/3 \sigma_{tot}$,
where $\sigma_{tot}$ is the total cross section. For a $p-$wave
interaction one has $\sigma_{tr}=2/5 \sigma_{tot}$. Therefore, the shear viscosity of a $p-$wave interaction is a factor 5/3 larger than the isotropic scattering.}, and
3) we implement the same scattering amplitude squared from SMASH, but multiplied by a factor $6/9$. This is due to the fact that
in \cite{Torres-Rincon:2012sda} an average scattering amplitude for all possible 9 initial states
$(\pi^\pm,\pi^0) \otimes (\pi^\pm,\pi^0)$ was used, whereas in the simulation we consider only 6 of these combinations (more specifically, scatterings
between pions of the same charge are not possible, if including only the $\rho$ meson).

\begin{figure}
 \includegraphics[width=85mm]{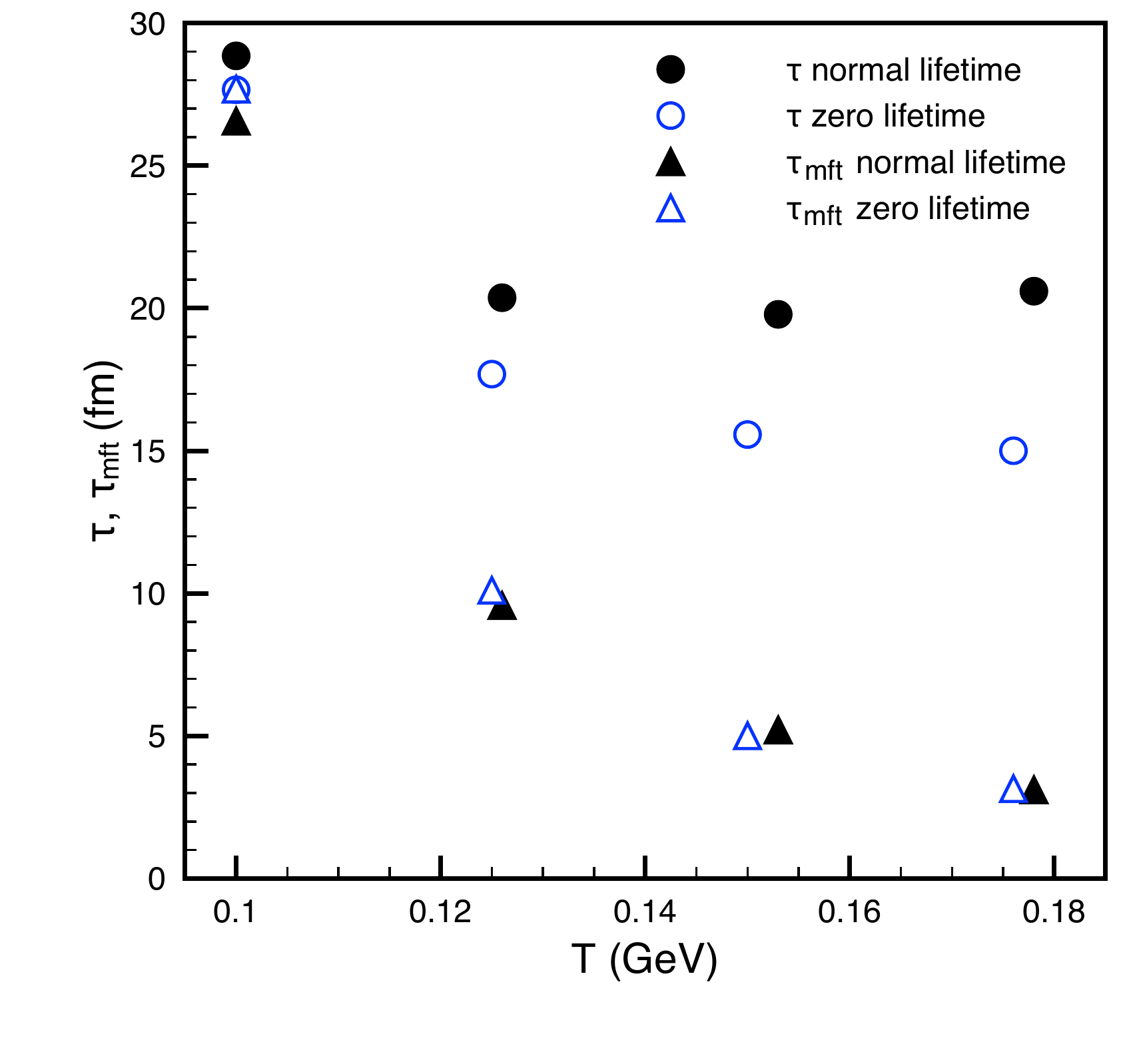}
\caption{Relaxation time $\tau$ and mean-free time $\tau_{mft}$ versus temperature in a $\pi-\rho$ gas, for different lifetimes.}
\label{pirho_tau}
\end{figure}

Figure \ref{pirho_viscosity} shows the effect of this difference on viscosities, as well
as the effect of forcing resonances to decay immediately in our transport model, which
effectively sets the lifetime of the $\rho$ resonance to zero and makes interactions
point-like. This has the effect of bringing the two results much closer together, to the point
where the two calculations are once again in strong agreement.

As shown in Fig.~\ref{pirho_tau}, the lowering of the shear viscosity when setting the
resonance lifetimes to zero is explained by looking at the relaxation time of the system
in both cases. As one can easily see, the relaxation time appears to be increasingly
reduced as one goes to higher temperatures; this suggests that the lifetime of resonances
can have a large impact on the relaxation time when the lifetime is not negligible with
respect to the mean free time of the particles in the system. In the latter case notice
that $\tau$ reaches a plateau at high temperatures. Intuitively, the finite lifetime of the $\rho$ meson delays the momentum transfer and therefore affects the relaxation dynamics. Note as well that the time between pion collisions
(or mean free time) $\tau_{mft}$ remains unaffected by this change in lifetimes.

\subsection{\label{hadron_gas_results}Full hadron gas}

\begin{figure}
 \includegraphics[width=85mm]{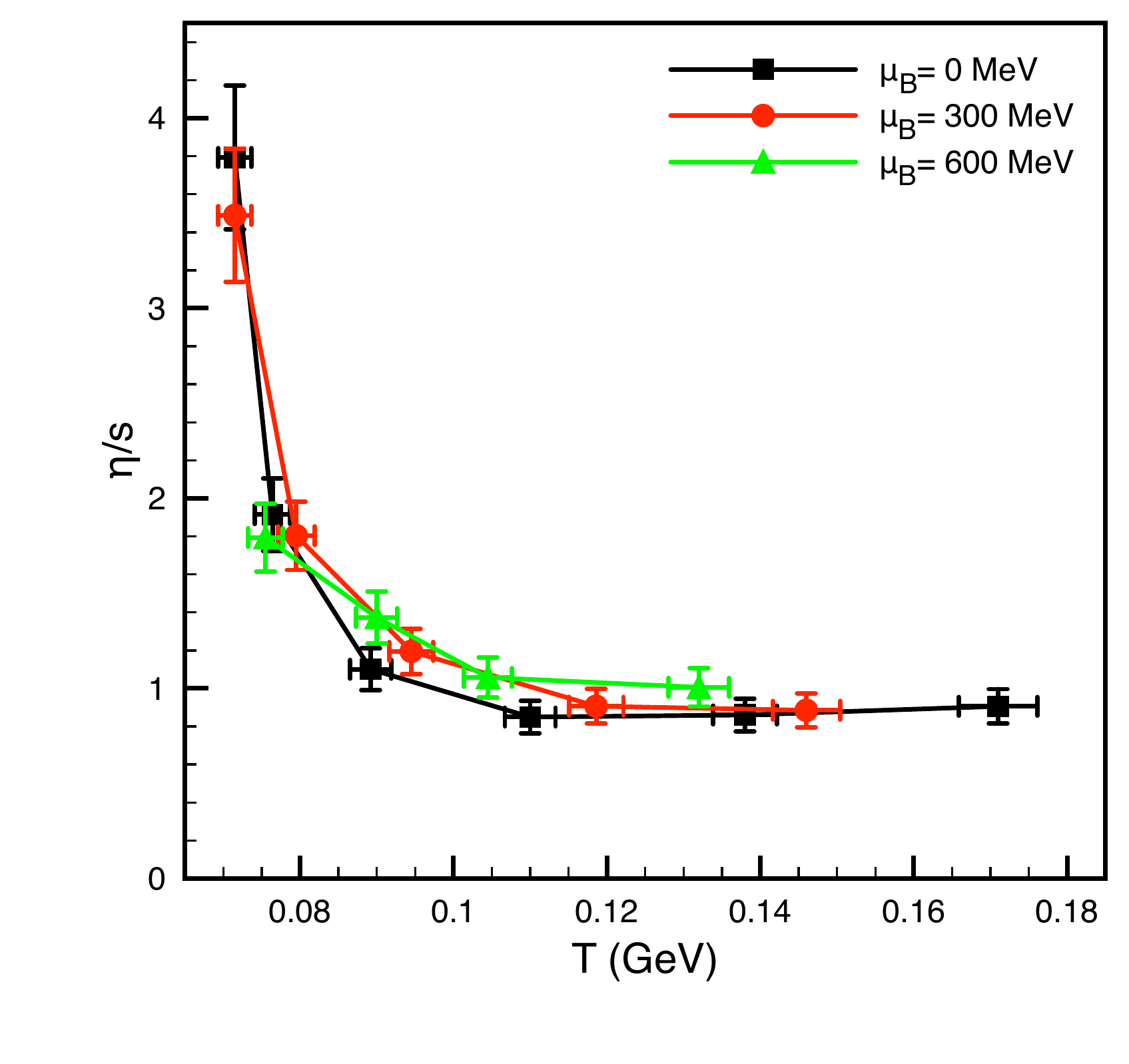}
\hspace{0mm}
 \includegraphics[width=85mm]{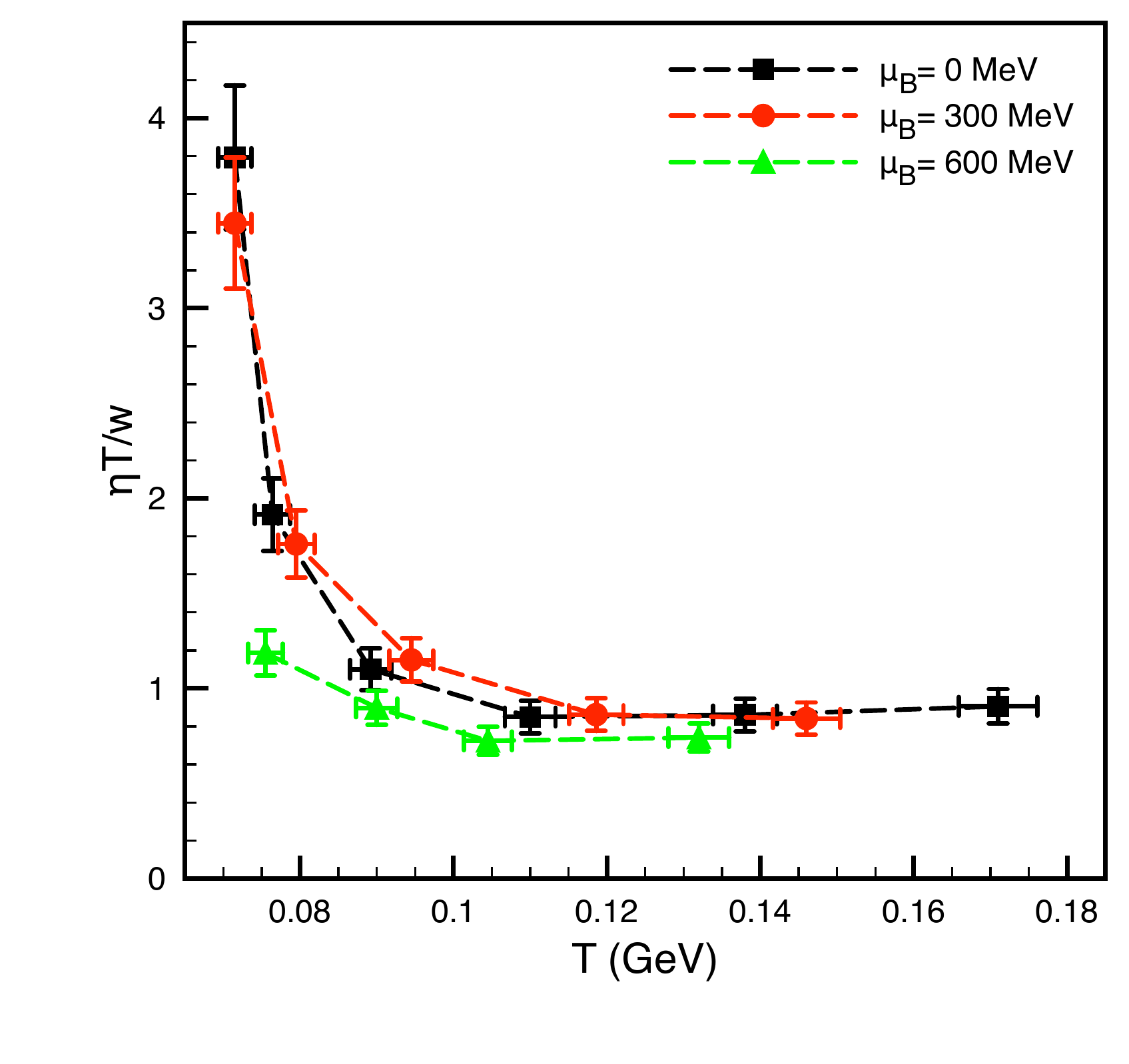}
\caption{Shear viscosity to entropy (top) and enthalpy (bottom) density ratios vs temperature, for various baryochemical potentials.}
\label{eta_over_s_vs_T}
\end{figure}

We now proceed to calculate the shear viscosity of a hadron gas as
simulated in the SMASH transport approach. Figure~\ref{eta_over_s_vs_T} shows both the
ratio of shear viscosity to entropy density and the ratio of shear viscosity to enthalpy
density. Although the former is used
as an input to hydrodynamic simulations~\cite{Romatschke:2007mq,Dusling:2007gi}, it has been argued that the latter provides more insight into the transport
properties of dense hadronic matter as this combination appears in the sound attenuation
length~\cite{Teaney:2003kp}. Here, both ratios are displayed. If we first
look at the zero baryonic chemical potential curves (which are identically the same, as expected), we
see that they display a decreasing profile at low temperatures, which corresponds to the
expected behavior of a liquid approaching a phase transition~\cite{Evans}. One also
notices that the shear viscosity to entropy/enthalpy density ratio reaches a plateau
around a temperature of $110$ MeV, and stays flat until around $170$ MeV, that is, for the
whole region around the temperature of $155$ MeV at which the phase transition is situated~\cite{Aoki:2006we}. The ratios start to increase slowly at temperatures
higher than $170$ MeV, but this is also the temperature above which quark and gluon degrees of freedom are becoming important. In SMASH, the cross-sections via resonance
excitation decrease at high energies and therefore our calculation is only meaningful in the hadronic region of the phase diagram. 

\begin{figure}
 \includegraphics[width=85mm]{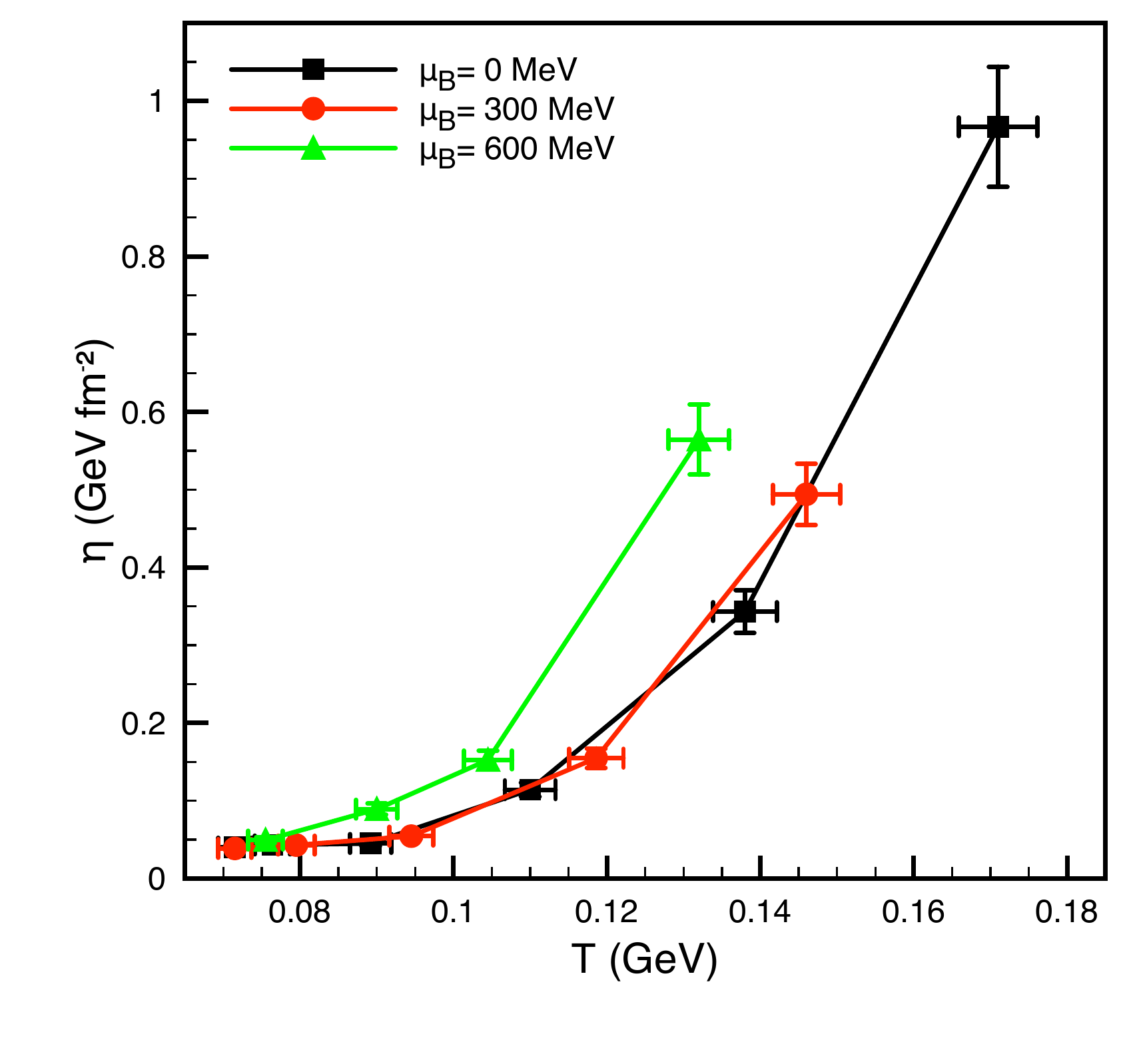}
\hspace{0mm}
 \includegraphics[width=85mm]{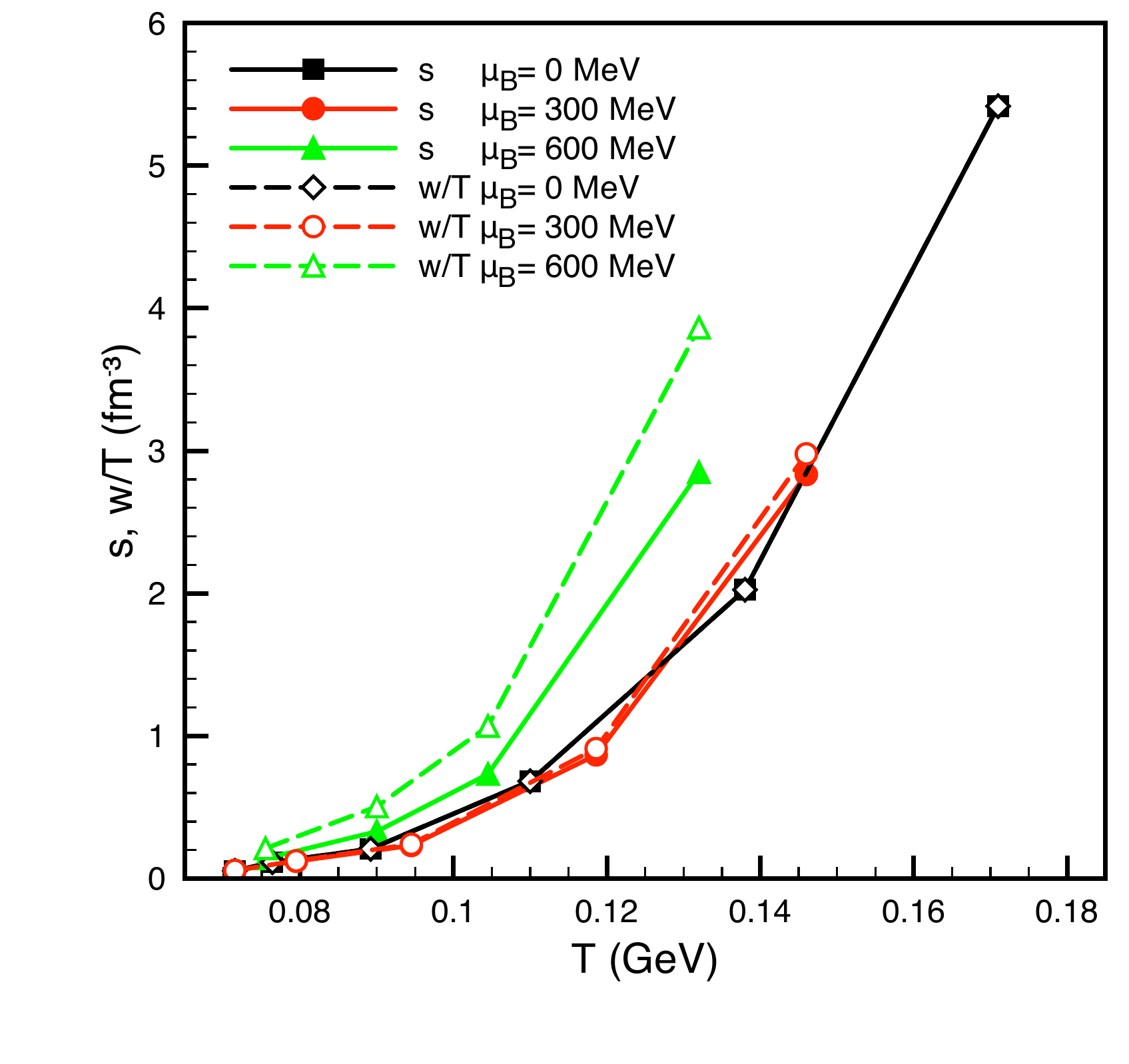}
\caption{Shear viscosity (top), entropy and enthalpy densities (bottom) vs temperature, for various baryonic chemical potentials.}
\label{eta_and_s_vs_T}
\end{figure}

Moving on to non-zero net baryon chemical potential, it appears that the ratio of shear
viscosity to entropy density is relatively independent of $\mu_B$ at every plotted
temperature, at least until values of the chemical potential of approximately 600-650 MeV. On the contrary, the ratio of shear viscosity to enthalpy density displays a difference when going to higher chemical potential. The difference between the two ratios highlights that the inclusion of the baryonic chemical
potential term in the entropy calculation (see Eq.~(\ref{gibbs_entropy})) can at times obscure some trends in the physical
picture.

For future reference and to help shed some light on the various features of
Fig.~\ref{eta_over_s_vs_T}, we now plot all components individually, namely the shear viscosity, entropy density and enthalpy density
(Fig.~\ref{eta_and_s_vs_T}). The top panel of Fig.~\ref{eta_and_s_vs_T} shows the behavior
of shear viscosity, which we find at all values of the chemical potential to be an
increasing function of temperature, as expected. Increasing chemical potential also
increases shear viscosity at equal temperature for all temperatures, which is also
expected.

The bottom panel of Fig.~\ref{eta_and_s_vs_T} simultaneously shows entropy and enthalpy
densities as functions of temperature. Since both of these quantities depend primarily on
the energy density of the system, it comes as no surprise that increasing the temperature
or baryon chemical potential leads to large increases here as well. Note here
that in this plot one sees very well the effect of including the baryonic chemical
potential in the entropy calculation, with the difference increasing from zero at $\mu_B=0$ MeV
to differences of 30 at 600 MeV. This can at least partly explain the shape of the corresponding curves in
Fig.~\ref{eta_over_s_vs_T}.

\begin{figure}
  \includegraphics[width=85mm]{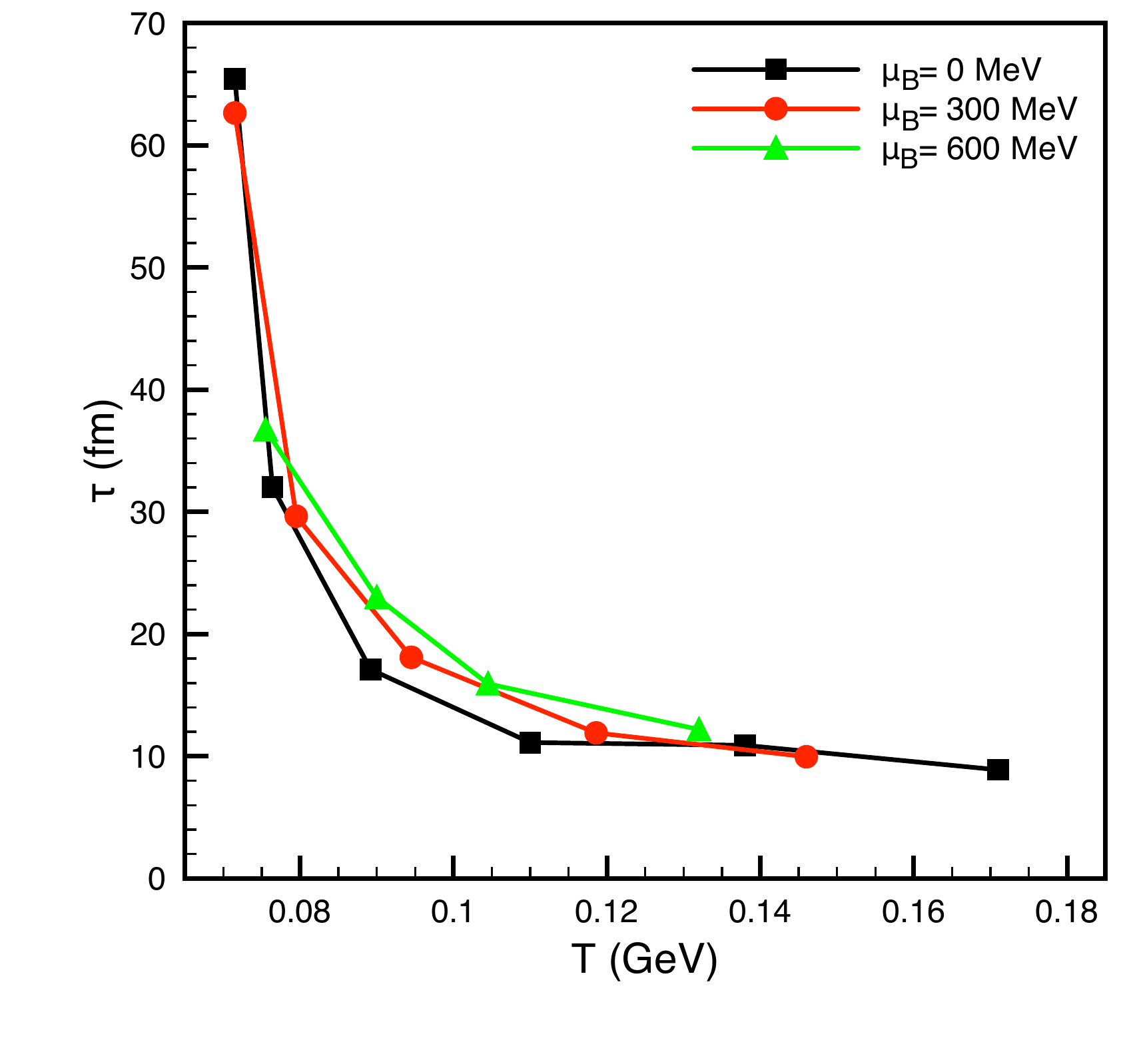}
  \caption{Shear relaxation time vs temperature, for various baryonic chemical potentials.}
  \label{tau_vs_T}
\end{figure}

In Fig.~\ref{tau_vs_T}, let us now further decompose the previous results by plotting
the shear relaxation time $\tau$, which comes into play in the calculation of the shear
viscosity (see Eq.~(\ref{final_shear_eq})). One should first note that the overall profile of these curves is
relatively similar to those of Fig.~\ref{eta_over_s_vs_T}.
This is expected, since as seen on Fig.~\ref{eta_and_s_vs_T}, $C^{xy} (0)$ rises
with the temperature in a way that is approximately matched by the rise in entropy
density/enthalpy, so that the final characteristic shape of $\eta/s$ or $\eta T/w$ is approximately mirroring the shape of the relaxation time.

At higher temperatures, there also
appears to be a trend of slightly increasing relaxation time as the baryonic chemical
potential increases. Since the composition of the gas is slowly moving towards baryonic
matter when increasing $\mu_B$, the baryonic resonances are becoming more
prevalent. If we consider that baryons generally have smaller cross-sections than their
mesonic counterparts, this easily explains the observed trend.

\begin{figure}
  \includegraphics[width=85mm]{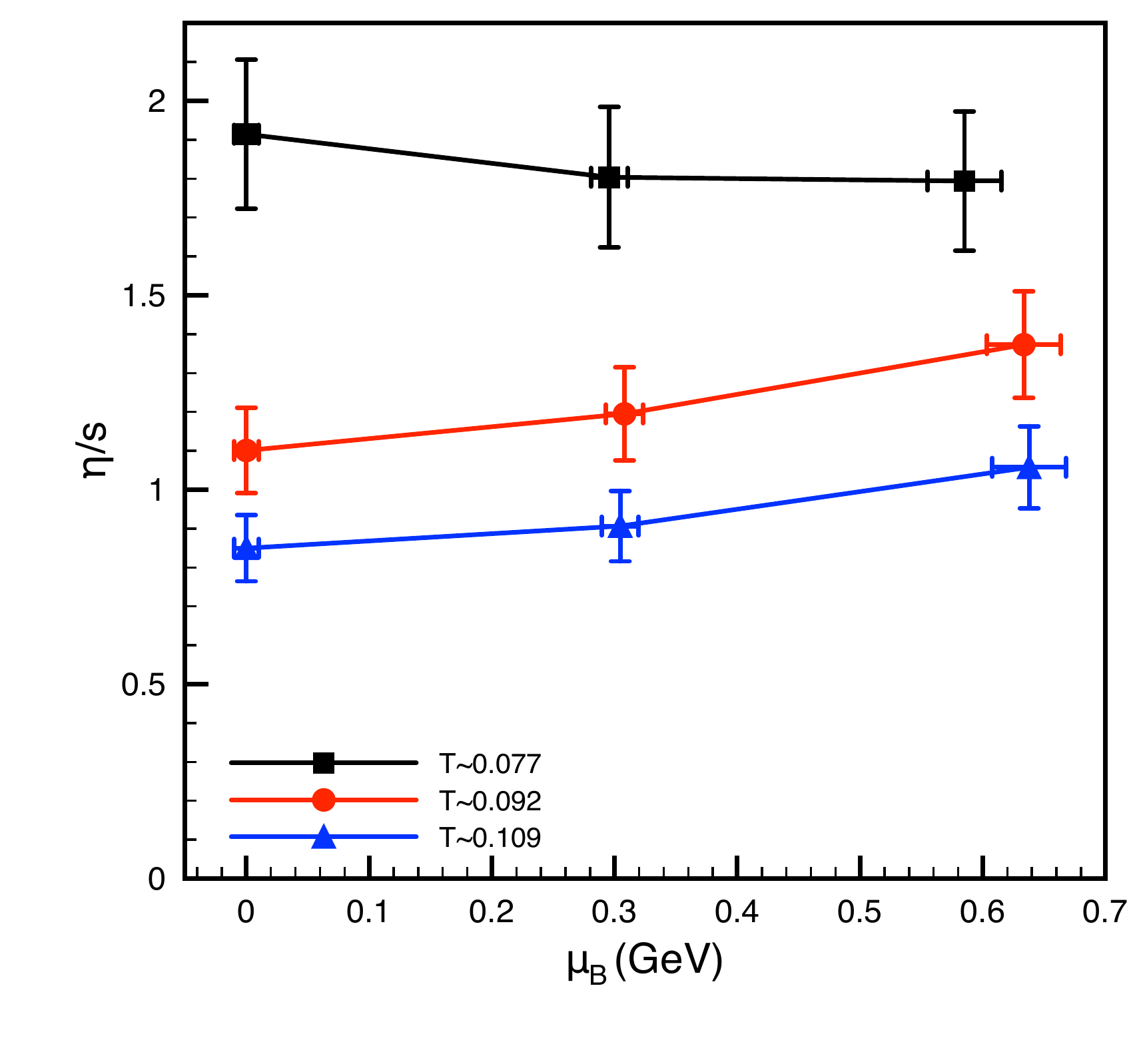}
  \caption{Shear viscosity to entropy density ratio vs baryonic chemical potential, for various temperatures.}
  \label{eta_s_vs_muB}
\end{figure}

Figure \ref{eta_s_vs_muB} shows the same data in a different way: instead of
taking temperature profiles at approximately constant baryonic chemical potential, the $\mu_B$ dependence of the shear viscosity to enthalpy
ratio is investigated at approximately constant temperature. As one can see, we observe for all
temperatures a slightly increasing plateau at these values of chemical potential; note that within error bars, this calculation is still consistent with no increase at all.
The calculated profile of the shear viscosity to entropy ratio at fixed temperature with
respect to the baryon chemical potential is actually quite close to what was computed in~\cite{Itakura:2007mx}. Notice that for the current range of temperatures and baryon chemical potentials, it has been
checked that the use of Fermi-Dirac instead of Boltzmann statistics has a negligible effect on
the observables.

\begin{figure}
  \includegraphics[width=85mm]{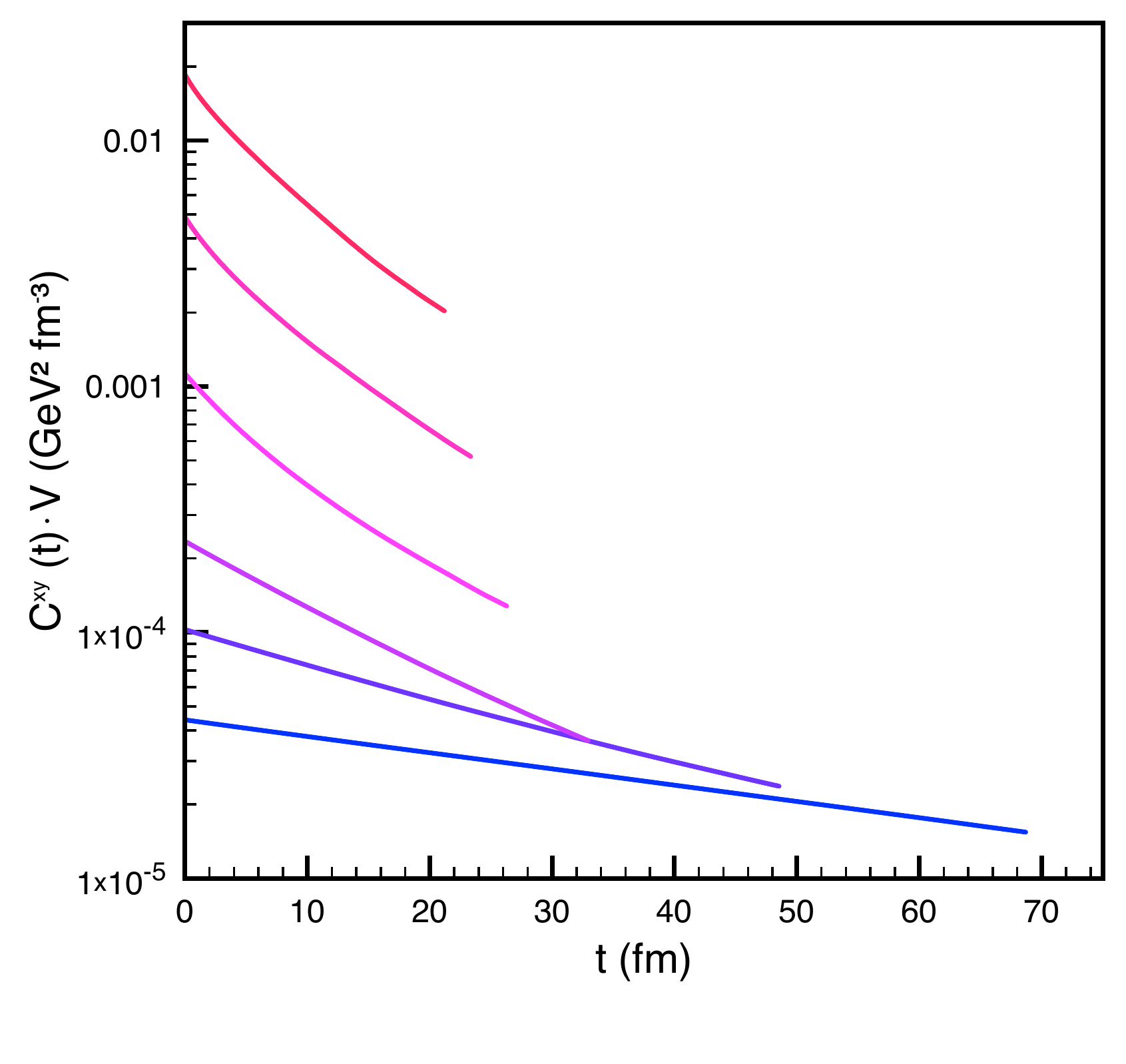}
  \caption{Typical volume-independent correlation functions for various temperatures at $\mu_B$ = 0. From blue to red, or bottom to top, the fitted corresponding temperatures are respectively 71.5, 76.4, 89.2, 110, 138 and 171 MeV. The plotting stops when the relative statistical error reaches 6\% in each case, and thus corresponds to the part of the curve which is fitted.}
  \label{correl_muB0}
\end{figure}

As a reference, we include some typical auto-correlation functions at $\mu_B$ = 0
(Fig.~\ref{correl_muB0}). As one can readily see, the slope of the function gets steeper
with rising temperature; this was directly visible from the previous Fig.~\ref{tau_vs_T},
where we saw the relaxation time (the inverse of the slope) steadily falling. The slightly
non-exponential behavior that one observes is investigated in more detail in \cite{Rose:2017ntg}.

\begin{figure*}
 \includegraphics[width=150mm]{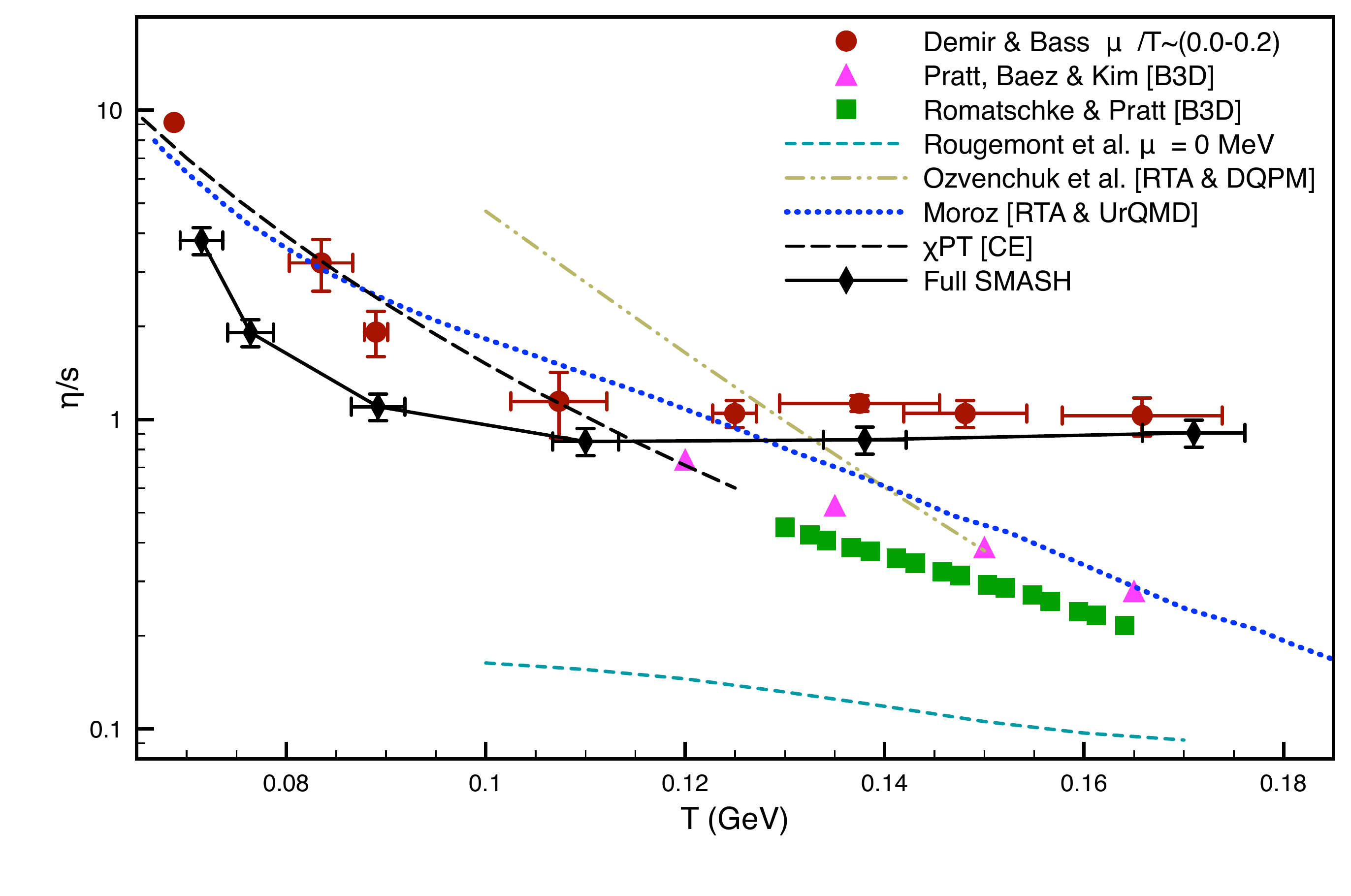}
\caption{Comparison of several calculations for the hadron gas $\eta/s$ at $\mu_B = 0$; see text for details.}
\label{comparison_0muB}
\end{figure*}

\subsection{\label{discussion}Discussion and comparison}

In this section, let us first summarize previous calculations of the shear viscosity over entropy density ratio of a hadron gas and then discuss in detail how they compare with our results. 
As mentioned earlier, the shear viscosity of the hadron gas is an active subject of
discussion, and multiple calculations of its value were performed
previously, especially for the zero baryon chemical potential case. A comparison of available calculations is presented in Fig.~\ref{comparison_0muB}. The Demir \& Bass~\cite{Demir:2008tr} calculation uses a similar
Green-Kubo formalism, but in the context of the UrQMD transport code~\cite{Bass:1998ca}. 
The Pratt, Baez \& Kim~\cite{Pratt:2016elw} curve is computed using the B3D code,
but this time by extracting the viscosity from Israel-Stewart equations, while obtaining the necessary other transport coefficients 
from the Kubo formalism. The Romatschke \& Pratt~\cite{Romatschke:2014gna} one uses once again the B3D cascade code, with the viscosity $\eta/s$
being assimilated directly to the response of the energy-momentum tensor to a velocity gradient. The Rougemont et al.
curve~\cite{Rougemont:2017tlu} is computed from a holographic correspondence using the Einstein-Maxwell-Dilaton model. Ozvenchuk et al. ~\cite{Ozvenchuk:2012kh} use
the relaxation time approximation for $\eta$ applied to the Parton-Hadron-String Dynamics approach~\cite{Cassing:2009vt}. 
Moroz~\cite{Moroz:2013vd} is an analytical calculation of the hadron gas shear viscosity using the relaxation time approximation and  modified UrQMD cross-sections (the EQCS2s set was used).
Finally, the $\chi$PT curve uses a Chapman-Enskog expansion to solve the Boltzmann equation relying on the lowest-order scattering
amplitude from chiral perturbation theory for the massive pion interaction~\cite{Torres-Rincon:2012sda}.

Let us know discuss the comparisons for each result starting from the low temperature region. 
Chiral perturbation theory~\cite{Gasser:1983yg} is the low-energy effective theory of QCD describing the dynamics of the
pseudo-Goldstone bosons, associated to the spontaneous symmetry breaking of the chiral symmetry. At lowest order in the chiral
expansion, the effective Lagrangian provides the scattering amplitude for the $\pi-\pi$ scattering~\cite{Scherer:2002tk}.

\begin{figure}
 \includegraphics[width=85mm]{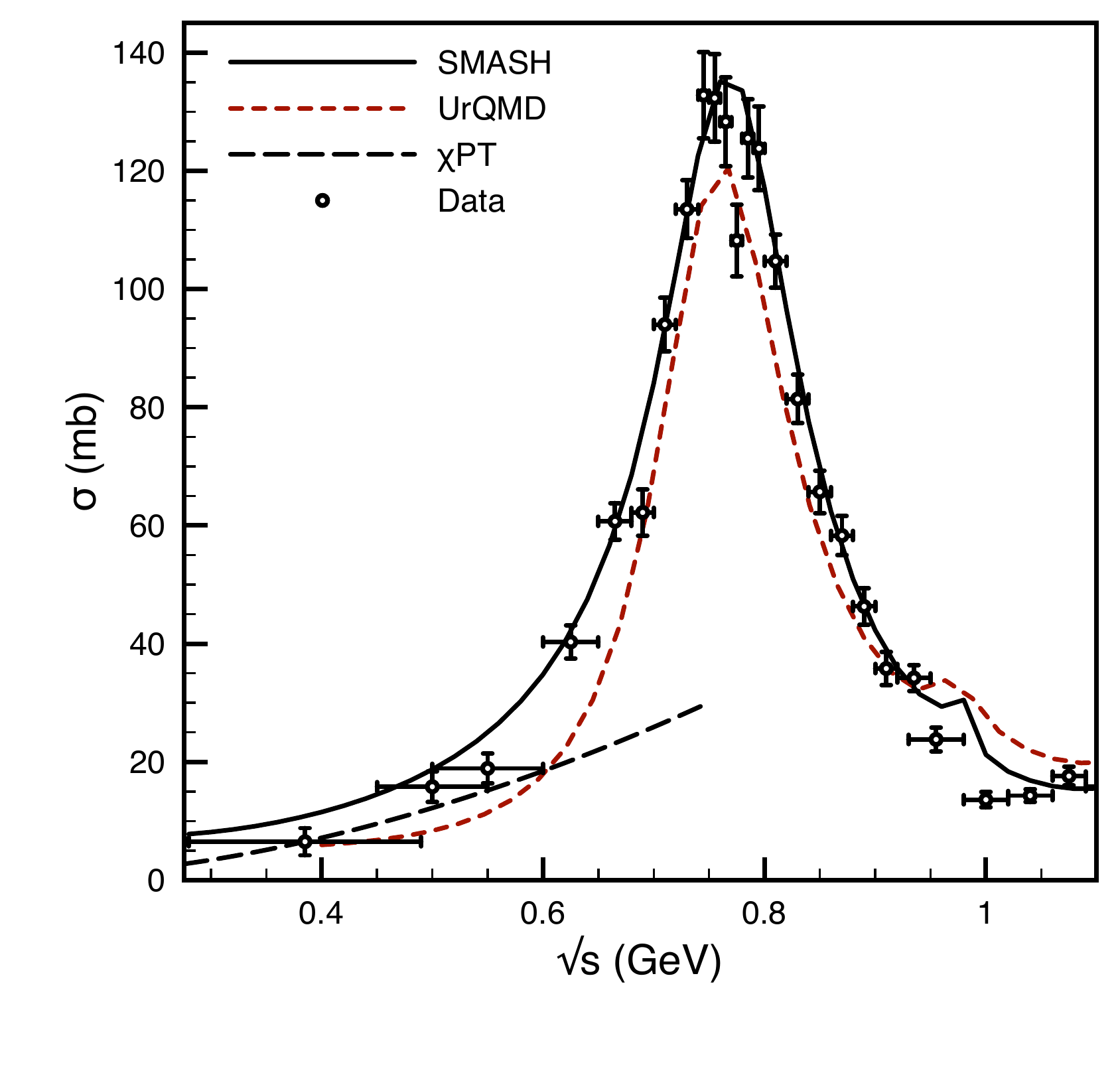}
\caption{Total $\pi^+ \pi^-$ cross-section in SMASH, UrQMD and LO $\chi$PT, compared to experimental data.}
\label{pipi_xsec}
\end{figure}

The LO $\chi$PT calculation with massive pions provides a model-independent reference value for $\eta/s$.
However, its validity is restricted to very low temperatures where the system can be approximated
to a gas of low-energy pions (up to $T\sim 70$ MeV~\cite{Itakura:2007mx}).
At low temperature, SMASH gives values of $\eta/s$ of the same order of magnitude, but one should not expect
a perfect matching to the LO $\chi$PT with our results: in the former, the $I=1$ channel
carries an angular dependent differential cross section (the lowest allowed partial wave is a $p$-wave)
as opposed to isotropic emissions in SMASH (as mentioned earlier, this accounts for differences of the order of up to 5/3 lower viscosity in pion dominated systems). Further differences exist between the transport
model and $\chi$PT calculations. As seen on Fig.~\ref{pipi_xsec}, the dominant $\pi^+ \pi^-$ cross section
at these low temperatures (which corresponds to the early part of the curve, $\sqrt{s} > 0.7-0.8$ GeV) is significantly larger in SMASH (especially when going to higher energies, where $\chi$PT cannot describe any kind of resonant interaction), which also contributes to lower viscosities. The $\chi$PT result additionally
contributes to the elastic $\pi^{\pm} \pi^{\pm}$ scattering, which is not taken into account in
the transport model, although it is expected this will have a more limited effect. Finally, in SMASH the pion scattering occurs by the formation of an intermediate resonance containing 
the inherent time delay due to its lifetime, which is not implemented in the $\chi$PT calculation.
This was mentioned in section~\ref{pirho} and will be explained in more detail later, although it should remain much smaller than the other effects at such small temperatures (see Fig.~\ref{pirho_viscosity}).

Differences between the SMASH and UrQMD description of $\pi\pi$ cross-sections can explain the low temperature discrepancy between our calculation and that of Demir \& Bass~\cite{Demir:2008tr}~
which used a comparable Green-Kubo method with this other transport code. Fig.~\ref{pipi_xsec} is useful
in this regard: although both transport approaches describe reasonably well the experimental data, at low energies 
SMASH tends to slightly overshoot it while UrQMD slightly undershoots it; this results in a cross-section that is sometimes twice as large in SMASH for this low energy. UrQMD also includes a flat 5mb elastic cross-section for
all $\pi\pi$ processes, which could have a slight cancellation effect (although this very small cross-section compared to the much larger inelastic one should not affect viscosity results so much). UrQMD similarly does not account for the $p$-wave nature of the $\rho$ resonance.

Therefore, an agreement between transport models such as SMASH or the conceptually similar UrQMD
and $\chi$PT remains unlikely, and it is then no real surprise that our results remain lower than this calculation at low temperature.
 As a corollary, although the result from Demir \& Bass appears to agree
very well with $\chi$PT, this can be accidental to a certain extent.

The Moroz calculation \cite{Moroz:2013vd} employs an
approach to calculate viscosity analytically from the relaxation time approximation in the
full hadron gas. The calculation uses a set of improved cross sections from the UrQMD model, including 
elastic plus quasielastic processes (EQCS2s set). The cross sections are implemented in analytical expressions
for the shear viscosity obtained from a Chapman-Enksog expansion of the Boltzmann equations. Although
information from resonances is encoded in the cross sections, the collision kernel only contains
elastic processes, and no retardation effects from finite lifetimes are considered. It matches quite
well the simpler $\chi PT$ expectation at low temperatures; our results however appear to 
remain consistently larger than this analytical curve for temperatures larger than 130 MeV.

In Ozvenchuk et al.~\cite{Ozvenchuk:2012kh} the relaxation time approximation is used to simplify
the Boltzmann equation and obtain a simple formula for the shear viscosity. Even when resonance formation is
implemented in PHSD simulations, the relaxation time is identified with the mean-free time extracted from 
collision rates in a box simulation. In this approach, the relaxation time contains no feedback from the resonance lifetimes.

Although exact values differ quite a bit, the general consensus appears to confirm the
expectation that the viscosity should generally decrease when approaching a phase
transition. That being said, two tendencies are appearing in this plot: some calculations
are constantly decreasing for the available data in this range of temperatures, and others appear to saturate at some point and form a plateau at higher temperature; our
calculation is among the latter.

Of note, the calculation by Demir \& Bass~\cite{Demir:2008tr} appears to have a similar behavior as
ours as temperature increases, the viscosity is also saturating at high
temperature. Even though our results are otherwise somewhat smaller, this similarity in
the behavior is striking when compared to the other tendency, which predicts a steadier
decrease to sometimes much lower values around the critical temperature. One of the common
points between UrQMD and SMASH is the treatment of interactions through
resonances, which have a non-zero lifetime as illustrated in
Fig.~\ref{lifetime_vs_point}. In contrast, almost all other calculations use point-like
interactions for a great portion of the considered hadronic interactions, if not all. 
The B3D transport code includes many long-lived resonances,
but simultaneously includes an overall constant cross-section of $\sigma=10$ mb~\cite{Novak:2013bqa},
which introduces many point-like interactions, and is thus somewhat hybridized in this regard\footnote{In principle, UrQMD also includes a point-like elastic cross-section extracted from the Additive Quark Model between all particles. However, on inspection, this elastic cross section turns out to be much smaller than the non-elastic resonance contribution (maximum of 1.35mb in the largest cases, on average almost an order of magnitude smaller). In consequence, only a small number of collisions are point-like in UrQMD.}. Rougemont et al.~\cite{Rougemont:2017tlu} use the completely different framework of holography,
and is therefore excluded from this categorization.

\begin{figure}
 \includegraphics[width=85mm]{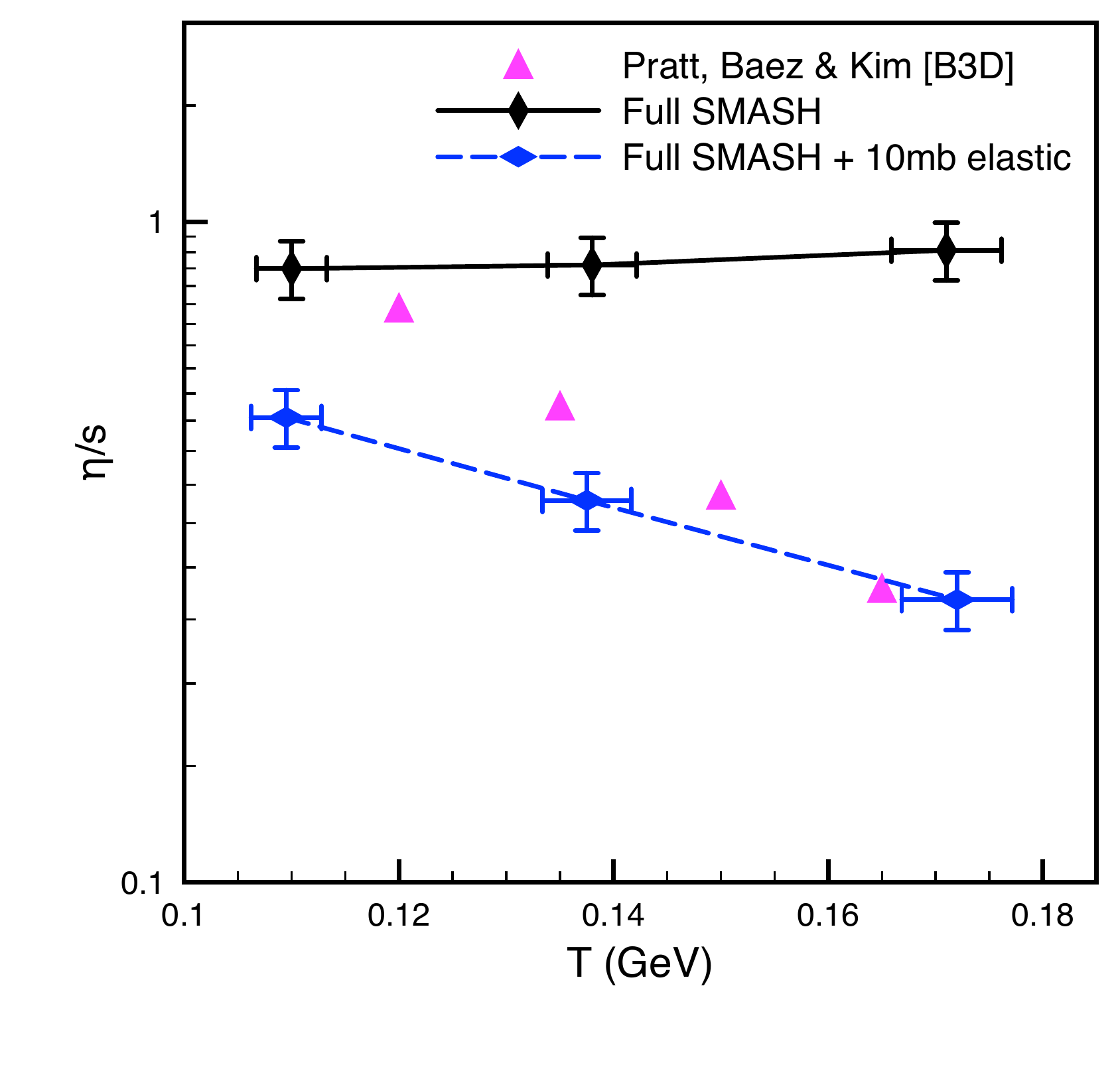}
\hspace{0mm}
 \includegraphics[width=85mm]{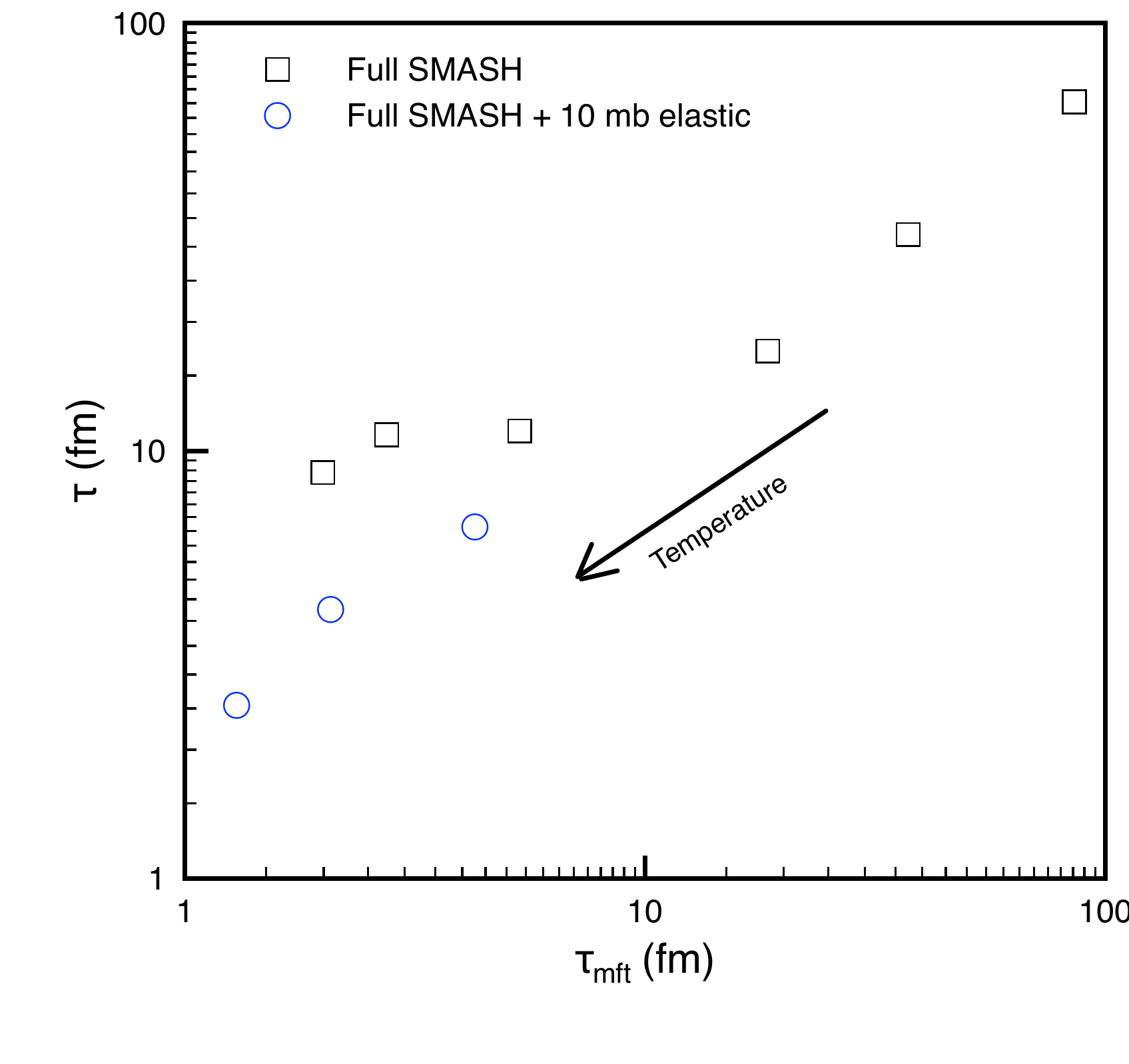}
\caption{Effect of adding a constant cross-section to all interactions on the viscosity (top) and on the characteristic times $\tau$ and $\tau_{mft}$ (bottom).}
\label{effect_of_elastic_xsec}
\end{figure}

To understand how resonance lifetimes affect the relaxation dynamics, consider a system without physically present resonances. The relaxation time is the characteristic time
in which the system approaches equilibrium after a slight departure from it. This time is
of microscopic origin, and it is assumed to be of the order of the collision time (or the inverse of
the scattering rate). Under a shear perturbation, particles with different momentum
will collide redistributing their energy to approach the thermal distribution. This
collision occurs on a time scale of the order of the mean free time, and therefore the relaxation time
should be of the same order. If the lifetime of the resonances is finite, but much smaller
than the mean free time, then the same picture holds, because the resonance will decay long
before the next collision is expected
to happen. Therefore, the transport process is unaffected by the generation of a resonance
if $\tau_{lifetime} \ll \tau_{mft}$. Again one expects that the relaxation time $\tau \sim
\tau_{mft}$. What happens to this picture if resonance lifetime is comparable (or larger) to
the mean free time? Then the transport process is blocked until the resonance eventually
decays, because it is only at that instant that the momentum exchange is finally
performed. The relaxation time is thus now limited by the lifetime of the resonance,
becoming independent of the $\tau_{mft}$, as we have checked numerically for the full
hadron box (bottom panel of Fig.~\ref{effect_of_elastic_xsec}).

This picture breaks down if a sufficient portion of the interactions are point-like.
If our explanation is correct, this breakdown in~\cite{Pratt:2016elw} is
caused by the large amount of elastic point-like collisions, which happen because of their
inclusion of a constant cross-section in B3D. To see if the
physical picture that we are depicting holds, let us also apply constant
isotropic cross-sections to all interactions in SMASH, so that a significant portion of
the collisions will now be point-like. The top panel of Fig.~\ref{effect_of_elastic_xsec}
shows the effect of such an adjustment, and we note two
differences. The first one is that all points are now at a lower value of shear viscosity,
which is explained by the increase in all cross-sections. The second difference
is more interesting, and concerns the profile of the curve: rather than saturating at a
given value, it now decreases constantly for this range of temperatures, which is what we
would expect from a system in which a large part of the interactions is now point-like, so
that the relaxation time is not affected by the lifetime of particles anymore.

For further evidence, let us look at the relation between the relaxation
time $\tau$ and the inverse of the scattering rate, the mean-free time
$\tau_{mft}$. In the case of no resonances~\cite{Pratt:2016elw} the
relaxation time increases linearly with the collision time, with a proportionality
factor of order 1. As seen on the bottom panel of Fig.~\ref{effect_of_elastic_xsec},
this expectation is fulfilled in SMASH for low temperatures, when the collision time of
particles is much larger than the lifetime of resonances. However, it breaks down at high
temperatures, when the collision time decreases to a value where the lifetime of
resonances is now large enough to impact the relaxation time of the system, thus forming a
plateau around $\tau \sim 10$ fm. When one includes a large number elastic
point-like collisions into SMASH, the plateau disappears and we recover a linear
dependency of order 1, even at high temperatures. This is once again in line with the
expectations of our resonance lifetime hypothesis.

\begin{figure}
  \includegraphics[width=85mm]{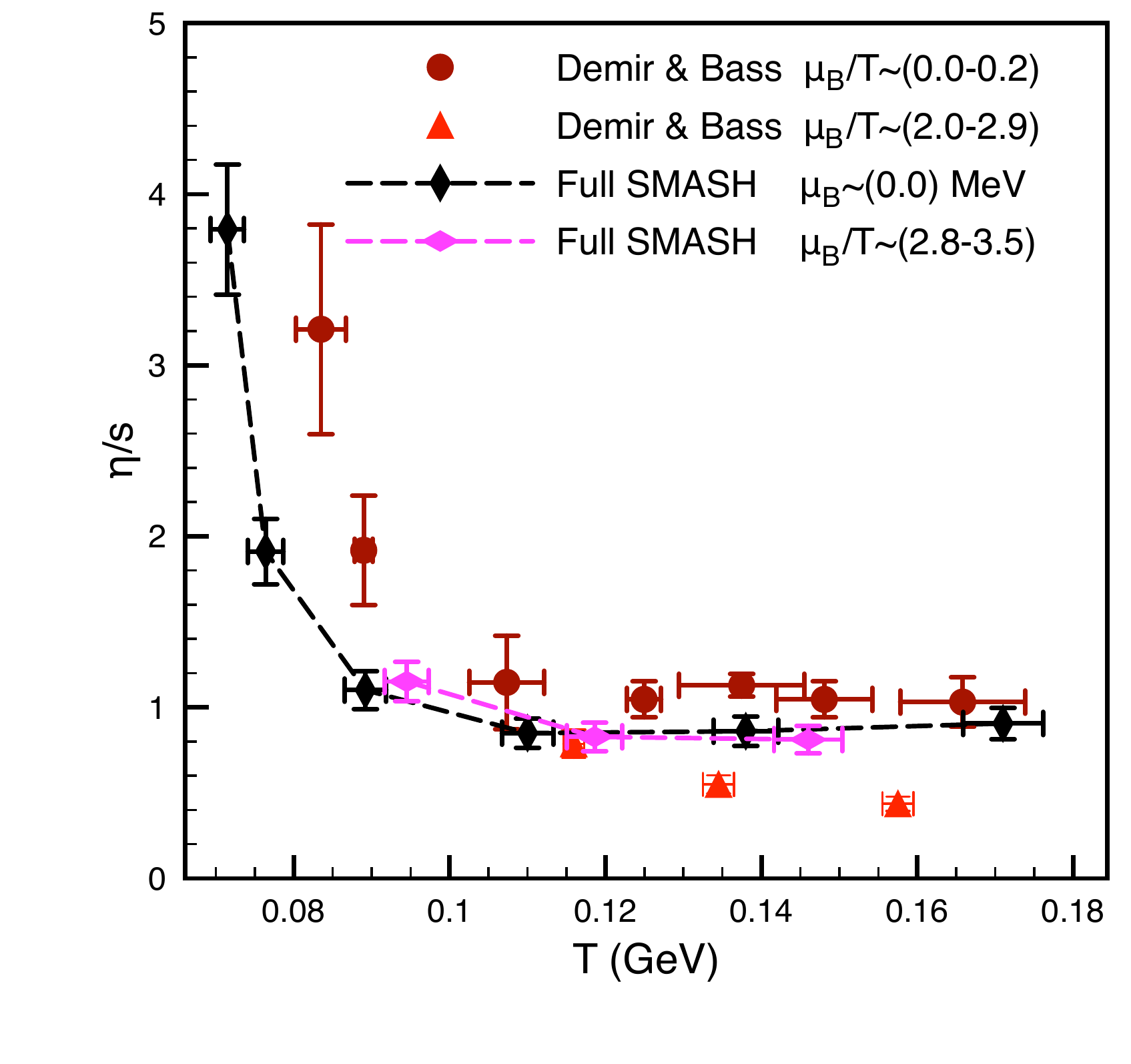}
  \hspace{0mm}
  \includegraphics[width=85mm]{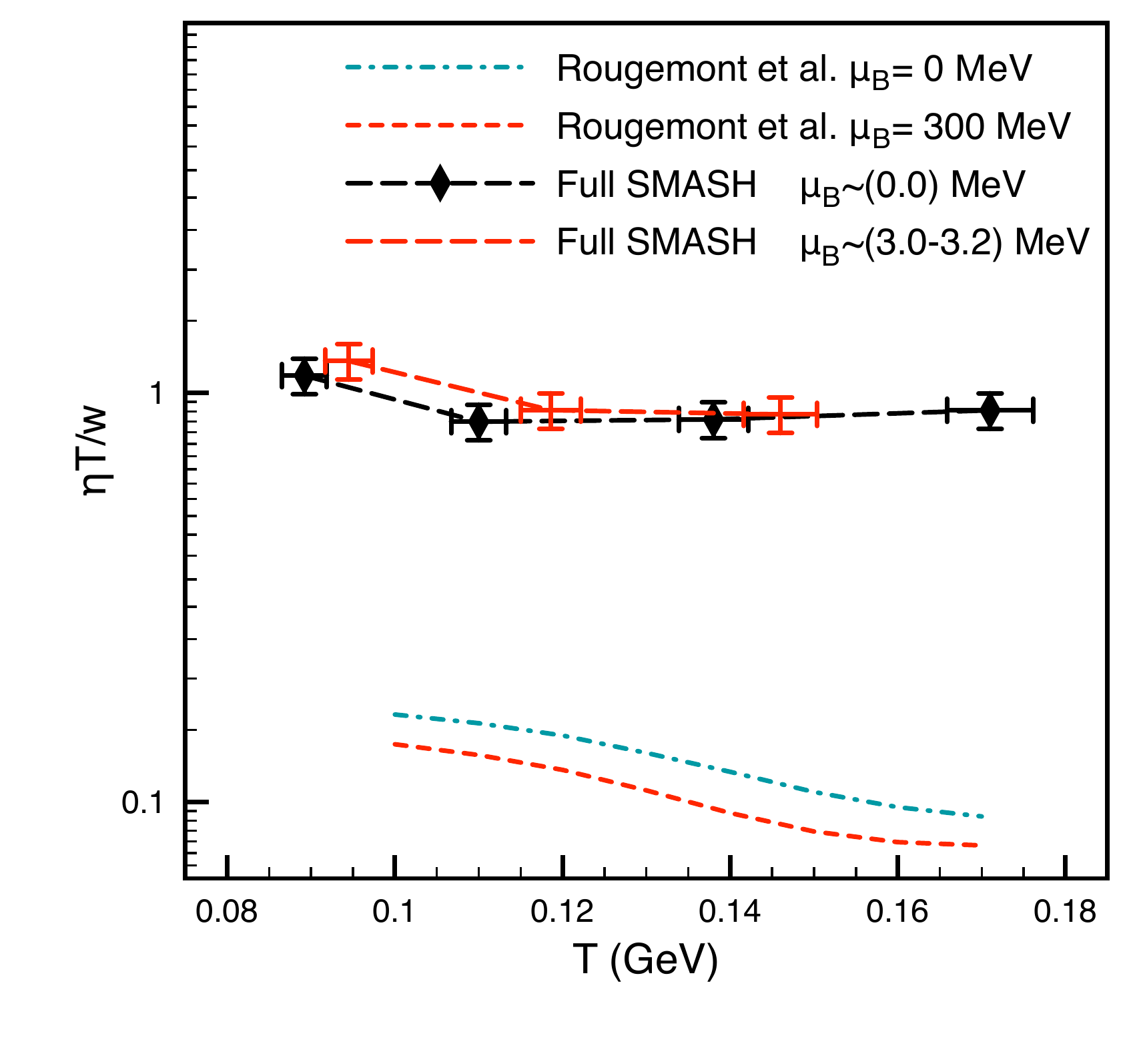}
  \caption{Non-zero chemical potential comparison with other models of shear viscosity. Comparison is possible with a previous Kubo calculation~\cite{Demir:2008tr} (top), and a result obtained from holography~\cite{Rougemont:2017tlu} (bottom)}
  \label{non-zero-muB-comparison}
\end{figure}

As a final remark, let us now considers the case of non-zero baryon chemical potential, where literature proves
to be a lot scarcer, although not inexistent. In this regard we present two comparisons
with other calculations (Fig.~\ref{non-zero-muB-comparison}). The first one was made with
the similar calculation from~\cite{Demir:2008tr} with UrQMD, and the second one from a holographic
approach~\cite{Rougemont:2017tlu}. In both cases, they observe a difference between
the zero and non-zero baryochemical potential results, with the non-zero case yielding a
smaller viscosity. In our calculation, both
cases are constant within errors. This discrepancy might be explained in the case
of Demir \& Bass by different methods of calculating the chemical potential term in the
entropy (which would also explain why the difference in their curves is growing with
temperature); a potential way of seeing whether this is a difference in the actual
models would be to compare the shear viscosity to enthalpy ratio instead. The approach in~\cite{Rougemont:2017tlu} is very different in nature, since it goes beyond the quasi-particle picture and and assumes strong coupling without confinement. Being close to the holographic result $\eta/s=1/4\pi$, it is natural that the resulting values of the shear viscosity to entropy ratio are smaller than in our approach. Still, it is interesting to note that the decrease for higher baryon chemical potentials is also observed just for lower values of the chemical potential than in our calculation, where the differences become significant only around $\mu_B = 600$ MeV. 

\section{\label{conclusion}Conclusion and Outlook}

After successfully testing the calculation of the viscosity in the recently developed
SMASH transport code in the simplest scenario of a single species interacting via constant
isotropic cross section, we presented a detailed analysis of the shear viscosity of a hadron
gas. The shear viscosity over entropy density coefficient was calculated for hadron matter
as a function of temperature and baryon chemical potential, and compared to several results in the literature.

The main conclusion of this study is that the details of the internal dynamics of the system
are of crucial importance when comparing transport coefficients in several approaches, as
microscopic details can be translated very differently into macroscopic effects. In
particular, the different treatment of the hadron-hadron interaction ($2 \rightarrow 2$
elastic collisions versus $2 \rightarrow 1 \rightarrow 2$ quasielastic dispersion) has
been seen to have large consequences from the point of view of transport, and the
calculation of the transport coefficients.

To turn this around, when more precise values for the temperature dependent viscosity are available from the extraction from experimental data, this could also be used to constrain the treatment of interactions in hadron transport approaches. Along these lines, one can argue that taking into account the medium modifications of the spectral functions of resonances usually results in a broadening and therefore a natural reduction of the lifetime, which counteracts the above described behavior. 

In the future, a more rigorous and mathematical analysis of the dependence of the
relaxation time on the resonance lifetime and mean free time might provide very
useful insights on questions pertaining to their interplay. Second, as previously noted,
almost all collisions in transport codes such as SMASH are treated isotropically;
as shown previously, the inclusion of more realistic angular distributions can in further investigations have an impact. At temperatures and baryon chemical potentials close to the phase transition, multi-particle interactions will also become relevant and it will be interesting to investigate their influence on the transport coefficients. 

\begin{acknowledgments}
We acknowledge Scott Pratt, Steffen Bass, Moritz Greif and Marcus Bleicher for useful discussions and Pierre Moreau for providing the ``Ozvenchuk et al.'' data in Fig.~\ref{comparison_0muB} from \cite{Ozvenchuk:2012kh}. This work was made possible thanks to funding from the Helmholtz Young Investigator Group VH-NG-822 from the Helmholtz Association and GSI, and supported by the Helmholtz International Center for the Facility for Antiproton and Ion Research (HIC for FAIR) within the framework of the Landes-Offensive zur Entwicklung Wissenschaftlich-Oekonomischer Exzellenz (LOEWE) program from the State of Hesse. Computing services were provided by the Center for Scientific Computing (CSC) of the Goethe University Frankfurt.
\end{acknowledgments}

\end{document}